\documentclass{ar2e}

\usepackage{latexsym} 
\usepackage{amssymb}  
\usepackage{amsfonts}
\usepackage{multirow} 
\usepackage{graphicx}

\begin{document}
\input epsf.def   
\input psfig.sty

\setcounter{topnumber}{2}
\renewcommand\topfraction{1.0}
\setcounter{bottomnumber}{1}
\renewcommand\bottomfraction{.3}
\setcounter{totalnumber}{3}
\renewcommand\textfraction{.0}
\renewcommand\floatpagefraction{.5}
\setcounter{dbltopnumber}{2}
\renewcommand\dbltopfraction{.7}
\renewcommand\dblfloatpagefraction{.5}

\setlength{\baselineskip}{14pt}
\renewcommand\baselinestretch{1}


\newcommand{\al}{\alpha}
\newcommand{\be}{\beta}
\newcommand{\ga}{\gamma}
\newcommand{\Ga}{\Gamma}
\newcommand{\de}{\delta}
\newcommand{\De}{\Delta}
\newcommand{\ep}{\varepsilon}
\newcommand{\eps}{\epsilon}
\newcommand{\ze}{\zeta}
\newcommand{\ka}{\kappa}
\newcommand{\la}{\lambda}
\newcommand{\La}{\Lambda}
\newcommand{\ph}{\varphi}
\newcommand{\del}{\nabla}
\newcommand{\si}{\sigma}
\newcommand{\Si}{\Sigma}
\renewcommand{\th}{\theta}   
\newcommand{\Up}{\Upsilon}
\newcommand{\om}{\omega}
\newcommand{\Om}{\Omega}
\newcommand{\imp}{~\Rightarrow}
\newcommand{\p}{\partial}
\newcommand{\<}{\langle} 
\renewcommand{\>}{\rangle} 
\newcommand{\ul}{\underline}
\newcommand{\txt}{\textstyle}
\newcommand{\dsp}{\displaystyle}
\newcommand{\h}{\hbox}
\newcommand{\ad}{\dagger}
\newcommand\eqn[1]{(\ref{#1})}      
\newcommand\Eqn[1]{Eq.~(\ref{#1})}  
\newcommand{\e}{ {\rm e} }
\newcommand{\beq}{\begin{equation}}
\newcommand{\eeq}{\end{equation}}
\newcommand{\ba}{\begin{array}}
\newcommand{\bea}{\begin{eqnarray}}
\newcommand{\ea}{\end{array}}
\newcommand{\eea}{\end{eqnarray}}
\newcommand{\bt}{\begin{tabular}}
\newcommand{\et}{\end{tabular}}

\newcommand{\lab}{\label}
\renewcommand{\slash}{\!\!\!\!/\,}
\newcommand{\sslash}{\!\!\!/\,}
\newcommand{\dotprod}{\!\cdot\!}
\newcommand{\vfilll}{\vskip 0pt plus 1filll}
\newcommand{\ol}{\overline}
\newcommand{\nl}{\hfil\break}
\newcommand{\ee}[1]{\times 10^{#1}}

\newcommand\hide[1]{}

\newcommand{\seq}{\!=\!}        

\newcommand{\half} {{\txt {1\over 2}}}
\newcommand{\third}{{\txt {1\over 3}}}
\newcommand{\quarter}{{\txt {1\over 4}}}
\newcommand{\sixth}{{\txt {1\over 6}}}
\newcommand{\eighth}{{\txt {1\over 8}}}

\newcommand{\back}{\mskip-.333\thinmuskip}
\newcommand{\phm}{\phantom{-}}

\renewcommand{\O}{ {\cal O} }
\newcommand{\eV}{{\rm eV}}
\newcommand{\keV}{{\rm keV}}
\newcommand{\MeV}{{\rm MeV}}
\newcommand{\GeV}{{\rm GeV}}
\newcommand{\fm}{{\rm fm}}
\newcommand{\Kelvin}{{\rm K}}
\newcommand{\Hz}{{\rm Hz}}

\newcommand{\psibar}{{\bar\psi}}

\newcommand{\vp}{{\bf p}}
\newcommand{\vk}{{\bf k}}
\newcommand{\vq}{{\bf q}} 

\newcommand{\dm}{\de \mu}
\newcommand{\ms}{m_s}
\newcommand{\Ms}{M_s}
\newcommand{\Minv}{{M^{-1}}}
\newcommand{\qslash}{{q\!\!\!/}}
\newcommand{\qvsq}{{\vec q}^{\,2}}
\newcommand{\Qt}{{\tilde Q}}
\newcommand{\Y}{{T_8}}

\newcommand{\percent}{\symbol{'045}}
\newcommand{\Msolar}{M_\odot} 
\newcommand{\Z}{\mathbb{Z}}


\jname{Annu. Rev. Nucl. Part. Sci.}
\jyear{2001}
\jvol{?}
\ARinfo{?}

\title{Color superconducting quark matter}

\markboth{Alford}{Color superconducting quark matter}

\author{Mark Alford
\affiliation{Dept.~of Physics and Astronomy, Glasgow University,
Glasgow G12 8QQ, UK}}


\begin{abstract}
I review recent progress
in our understanding of the color superconducting phase of 
matter above nuclear density, giving particular emphasis
to the effort to find observable signatures of the
presence of this phase in compact stars.
\end{abstract}

\maketitle

\section{INTRODUCTION}
\label{sec:int}

In the last 30 years, the idea that matter has subnuclear constituents
(``quarks'') has emerged from the
speculative edges of particle physics to become
a firmly established physical theory.
Quantum Chromodynamics (QCD),
the theory of the interaction of quarks via the gluon field,
is now one of the pillars of the
standard model. In high momentum processes, perturbative QCD has been
verified comprehensively.
For the spectrum and structural properties of the hadrons, the 
strongly-coupled intractability of the theory is giving way to
the brute-force numerical methods of lattice QCD, and again yields
agreement with experiment.
Even so, there remain tantalizing questions.  As well as 
predicting the properties and behavior of small numbers
of particles, QCD should also be able to tell us about the
thermodynamics of matter in the realm of unimaginably high temperatures
and densities at which it comes to dominate the physics.
However, only in the last few years have these regions begun to be probed
experimentally, and our theoretical understanding of them
remains elementary.

Progress has been fastest for matter at high temperature and low
density.  Lattice gauge calculations \cite{Karsch_Tc} show that chiral
symmetry is restored and deconfinement occurs at a temperature $T\sim
180~\MeV$.

At high densities, in contrast, lattice methods have
so far proved unhelpful. We are still trying to establish
the symmetries of the ground state, and find
effective theories for its lowest excitations.
These questions are of direct physical
relevance: an understanding of the symmetry properties
of dense matter can be expected to inform our understanding
of neutron star astrophysics and perhaps also heavy
ion collisions which achieve high baryon densities without
reaching very high temperatures.

In this review I will explore the progress that has been 
made in the last few years in understanding the possible phases
of QCD at low temperatures and high densities, 
and go on to discuss the possible observable signatures
in compact stars phenomenology.
For alternative explanations and emphasis, I refer the reader
to previous review articles on this topic~\cite{Reviews}.

\subsection{The Fermi surface and Cooper instability}

One of the most striking features of QCD is asymptotic freedom: the
force between quarks becomes arbitrarily weak as the characteristic
momentum scale of their interaction grows larger.  This immediately
suggests that at sufficiently high densities and low temperatures,
matter will consist of a Fermi sea of essentially free quarks,
whose behavior is dominated by the freest of them all: the
high-momentum quarks that live at the Fermi surface.

Such a picture is too naive. It was shown by Bardeen, Cooper, and
Schrieffer (BCS)~\cite{BCS} that in the presence of attractive interactions a
Fermi surface is unstable.  If there is {\em any} channel in which the
quark-quark interaction is attractive, then the true ground state of
the system will not be the naked Fermi surface, but rather a
complicated coherent state of particle and hole pairs---``Cooper
pairs''.

This can be seen intuitively as follows. Consider a system of free
particles.  The Helmholtz free energy is $F= E-\mu N$, where $E$ is
the total energy of the system, $\mu$ is the chemical potential, and
$N$ is the number of particles. The Fermi surface is defined by a
Fermi energy $E_F=\mu$, at which the free energy is minimized, so
adding or subtracting a single particle costs zero free energy.  Now,
suppose a weak attractive interaction is switched on.  BCS showed that
this
leads to a complete rearrangement of the states near the
Fermi surface, because it costs no free energy to
make a pair of particles (or holes), and the attractive
interaction makes it favorable to do so. Many such pairs will therefore
be created, in all the modes near the Fermi surface, and these pairs,
being bosonic, will form a condensate. The ground state will be a
superposition of states with all numbers of pairs, breaking the
fermion number symmetry. An arbitrarily weak interaction has lead to
spontaneous symmetry breaking.

In condensed matter systems, where the relevant fermions are electrons,
the necessary attractive interaction has been hard to find.
The dominant interaction
between electrons is the repulsive electrostatic force, but
in the right kind of crystal there are attractive phonon-mediated
interactions that can overcome it. In these materials
the BCS mechanism leads to
superconductivity, since it causes Cooper pairing of electrons, which
breaks the electromagnetic gauge symmetry, giving mass to the photon
and producing the Meissner effect (exclusion of magnetic fields from a
superconducting region). It is a rare and delicate state, easily
disrupted by thermal fluctuations, so superconductivity
only survives at low temperatures.

In QCD, by contrast, the dominant gauge-boson-mediated interaction
between quarks is itself attractive
\cite{Barrois, BarroisPhD,BailinLove,ARW2,RappETC}.
The relevant degrees of
freedom are those which involve quarks with momenta near the Fermi
surface. These interact via gluons, in a manner described by QCD. The
quark-quark interaction has two color channels available, the antisymmetric
$\bar{\bf 3}$, and the symmetric {\bf 6}. It is attractive in the
$\bar{\bf 3}_A$: this can be seen in single-gluon-exchange or by
counting of strings.

Since pairs of quarks cannot be color singlets,
the resulting condensate will break the local color symmetry
$SU(3)_{\rm color}$.  We call this ``color superconductivity''.
Note that the quark pairs play the same role here as the Higgs particle
does in the standard model: the color-superconducting phase
can be thought of as the Higgsed (as opposed to confined)
phase of QCD.

It is important to remember that the breaking of a gauge symmetry
cannot be characterized by a gauge-invariant local order parameter
which vanishes on one side of a phase boundary. The superconducting
phase can be characterized rigorously only by its global symmetries.
In electromagnetism there is a non-local order parameter, the 
mass of the magnetic photons, that corresponds physically to the Meissner
effect and distinguishes the free phase from the superconducting one.
In QCD there is no free phase: even without pairing the gluons are not
states in the spectrum.  No order parameter distinguishes the Higgsed
phase from a confined phase or a plasma, so we have to look at the
global symmetries.

In most of this paper we will take an approach similar to that used
in analyzing the symmetry breaking of the standard model, and
discuss the phases of dense QCD in terms of a gauge-variant
observable, the diquark condensate, which is analogous
to the vacuum expectation value (VEV) of the Higgs field. 
However, this is only a convenience, and we will
be careful to label different phases by their unbroken global
symmetries, so that they can always be distinguished by
gauge-invariant order parameters.

\subsection{The gap equation}
\label{sec:gap}

To decide whether or not fermions condense in the ground state, one
can explicitly construct a wavefunctional with the appropriate
pairing, and use a many-body variational approach.
But the field-theoretical
approach, though less concrete, is more general, and I will briefly
describe it here.

The important quantity is the quark self energy, i.e.~the one-particle
irreducible (1PI) Green function of two quark fields.
Its poles will give the gauge-invariant masses of the
quasiquarks, the lowest energy fermionic excitations around
the quark Fermi surface.
To see if condensation (chiral condensation,
flavor-singlet quark pairing, or whatever)
occurs in some channel, one writes down a
self-consistency equation, the ``gap equation'', for a self energy with
that structure, and solves it to find the actual self energy (the
gap). If it is zero, there is no condensation in that channel. If not,
there can be condensation, but it may just be a local minimum of the
free energy.  There may be other solutions to the gap equation, and
the one with the lowest free energy is the true ground state.

\begin{figure}
\begin{center}
 \psfig{figure=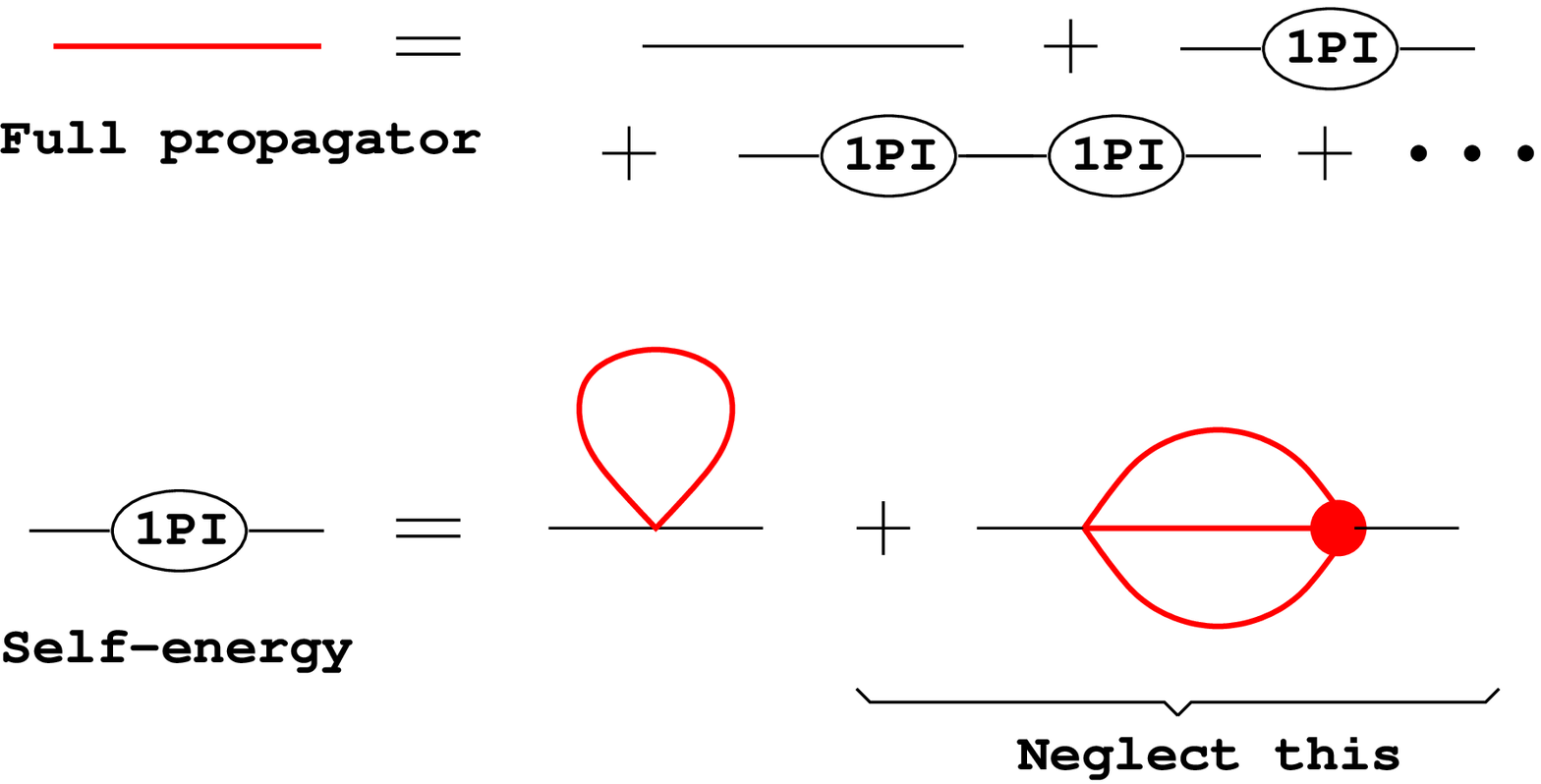,width=3.5in}
\end{center}
\caption{Mean-field Schwinger-Dyson (gap) equations}
\label{fig:SD}
\end{figure}

There are several possible choices for the interaction to be used
in the gap equation. At asymptotically high densities
QCD is weakly coupled, so one gluon exchange is appropriate. 
Such calculations 
\cite{Son,SW-pert,PR-pert,Hong-pert,PR-Tc,HMSW,rockefeller,Hsu2,SchaeferPatterns,ShovWij}
are extremely important, since they demonstrate
from first principles that color superconductivity
occurs in QCD.
However, the density regime of physical interest for neutron stars
or heavy ion collisions is 
up to a few times nuclear density ($\mu \lesssim 500~\MeV$)
and weak coupling calculations are unlikely to
be trustworthy in that regime. In fact, 
current weak-coupling calculations cannot be extrapolated
below about $10^8~\MeV$ because of gauge dependence arising from
the neglect of vertex corrections~\cite{RajagopalShuster}.
There have also been some preliminary investigations of confinement-related
physics such as a gluon condensate \cite{AKS,EKT}.

The alternative is to use some phenomenological interaction that
can be argued to capture the essential physics of QCD in the
regime of interest. The interaction can be normalized to reproduce
known low-density physics such as the chiral condensate, and
then extrapolated to the desired chemical potential.
In two-flavor theories, the instanton vertex is a natural choice
\cite{ARW2,RappETC,BergesRajagopal,CarterDiakonov},
since it is a four-fermion interaction. With more flavors,
the one gluon exchange vertex without a gluon propagator
\cite{BailinLove,IwaIwa,ARW3}
is more convenient.
It has been found that these both give the same results,
to within a factor of about 2. This is well within the inherent 
uncertainties of such phenomenological approaches.
In the rest of this paper we will therefore not always
be specific about the exact interaction used to obtain
a given result.
One caveat to bear in mind is that the
single-gluon-exchange interaction is symmetric under $U(1)_A$, and so
it sees no distinction between condensates of the form $\<q C q \>$
and $\<q C \ga_5 q \>$.  However, once instantons are included the
Lorentz scalar $\<q C \ga_5 q \>$ is favored,~\cite{ARW2,RappETC} so 
in single-gluon-exchange calculations the parity-violating condensate
is usually ignored.

The mean-field approximation to the Schwinger-Dyson equations
is shown diagramatically in Figure~\ref{fig:SD}, relating
the full propagator to the self-energy. In the mean-field
approximation, only daisy-type diagrams are included in the resummation,
vertex corrections are excluded.
Algebraically, the equation takes the form
\beq\label{gap:SD}
\Si(k) = -{1\over (2\pi)^4} \int \!d^4q\, \Minv(q) D(k-q),
\eeq
where $\Si(k)$ is the self-energy,
$M$ is the full fermion matrix (inverse full propagator),
and $D(k-q)$ is the vertex,
which in NJL models will be momentum-independent,  but 
in a weak-coupling QCD
calculation will include the gluon propagator and couplings.
Since we want to study quark-quark condensation,
we have to write propagators in
a form that allows for this possibility, just as to study chiral symmetry
breaking it is necessary to use 4-component Dirac spinors rather than
2-component Weyl spinors, even if there is no mass term in the action.
We therefore use Nambu-Gorkov
8-component spinors, $\Psi = (\psi,\psibar^T)$,
so the self-energy $\Si$ can include a quark-quark pairing
term $\De$. The fermion matrix $M$ then takes the form
\beq\label{gap:ferm}
M(q) = M_{\rm free} + \Si =
\left(\ba{cc} \qslash + \mu\ga_0 & \ga_0\De\ga_0 \\ 
 \De & (\qslash -\mu\ga_0)^T \ea \right).
\eeq
Equations \eqn{gap:SD} and \eqn{gap:ferm} can be combined to
give a self-consistency condition for $\De$, the gap equation.
If the interaction is a point-like four-fermion NJL interaction
then the gap parameter $\De$ will be a color-flavor-spin matrix,
indepndent of momentum. If the gluon propagator is included, $\De$
will be momentum-dependent, complicating the analysis considerably.

In NJL models, the simplicity of the model has allowed
renormalization group analyses~\cite{SW-RG,EHS}
that include a large class of four-fermion interations, and follow
their running couplings as modes are integrated out.
This confirms that in QCD with two and three massless quarks 
the most attractive
channels for condensation are those corresponding to the
``2SC'' and ``CFL'' phases studied below.
Calculations using random matrices, which represent very
generic systems, also show that diquark condensation
is favored at high density~\cite{randmat}.

Following through the analysis outlined above, 
one typically finds gap equations
of the form
\beq
1 = K \int_0^\La k^2dk\, \frac{1}{\sqrt{(k-\mu)^2 + \De^2}},
\eeq
where $K$ is the NJL four-fermion coupling.
In the limit of small gap, the integral can be
evaluated, giving 
\beq
\De \sim \La\exp\Bigl(\frac{\rm const}{K\mu^2}\Bigr).
\label{gap:NJL}
\eeq
This shows the non-analytic dependence of the gap on the 
coupling $K$. Condensation is a nonpertubative effect that
cannot be seen to any order in perturbation theory.
The reason it can be seen in the diagrammatic Schwinger-Dyson
approach is that there is an additional ingredient: 
an ansatz for the form of the self energy. This corresponds to
guessing the from of the ground state wavefunction in a many-body
variational approach. All solutions to gap equations therefore
represent possible stable ground states, but to find the favored
ground state their free energies must be compared, and even then
one can never be sure that the true ground state has been found,
since there is always the possibility that there is another
vacuum that solves its own gap equation and has an even lower free energy.

In weak-coupling QCD calculations, where the full
single-gluon-exchange vertex complete with gluon propagator
is used, the gap equation takes the form~\cite{Barrois,BarroisPhD,Son,PR-Tc}
\beq
\Delta \sim \mu 
\frac{1}{g^5}
\exp\Bigl(-\frac{3\pi^2}{\sqrt{2}}
\frac{1}{g}\Bigr)\ ,
\label{gap:QCD}
\eeq
or, making the weak-coupling expansion in the QCD gauge coupling
$g$ more explicit,
\beq
\ln\Bigl(\frac{\De}{\mu}\Bigr) = -\frac{3\pi^2}{\sqrt{2}}
\frac{1}{g} - 5 \ln g + {\rm const} + \O(g).
\eeq
This gap equation has two interesting features.
Firstly, it does not correspond to what you would naively expect
from the NJL model of single gluon exchange, in which the
gluon propagator is discarded and $K\propto g^2$,
yielding $\De \sim \exp(-1/g^2)$. The reason~\cite{BarroisPhD,Son} is
that at high density the gluon propagator has an infrared divergence
at very small angle scattering, since magnetic gluons are
only Landau damped, not screened. This
divergence is regulated by the gap itself, weakening its dependence
on the coupling.

Secondly, in \eqn{gap:QCD} we have left unspecified the energy scale
at which the coupling $g$ is to be evaluated.
Natural guesses would be $\mu$ or $\De$.
If we use $g(\mu)$ and assume it runs according to the one-loop
formula $1/g^2 \sim \ln \mu$ then the exponential factor in
\eqn{gap:QCD} gives very weak suppression, and is in fact
overwhelmed by the initial factor $\mu$, so that the gap
rises without limit at asymptotically high density,
although $\De/\mu$ shrinks to zero so that weak-coupling methods
are still self-consistent.
This means that color superconductivity will inevitably
dominate the physics at high enough densities.

\section{TWO MASSLESS QUARK FLAVORS}
\label{sec:2flavor}

\begin{figure}[htb]
\centerline{
 \psfig{figure=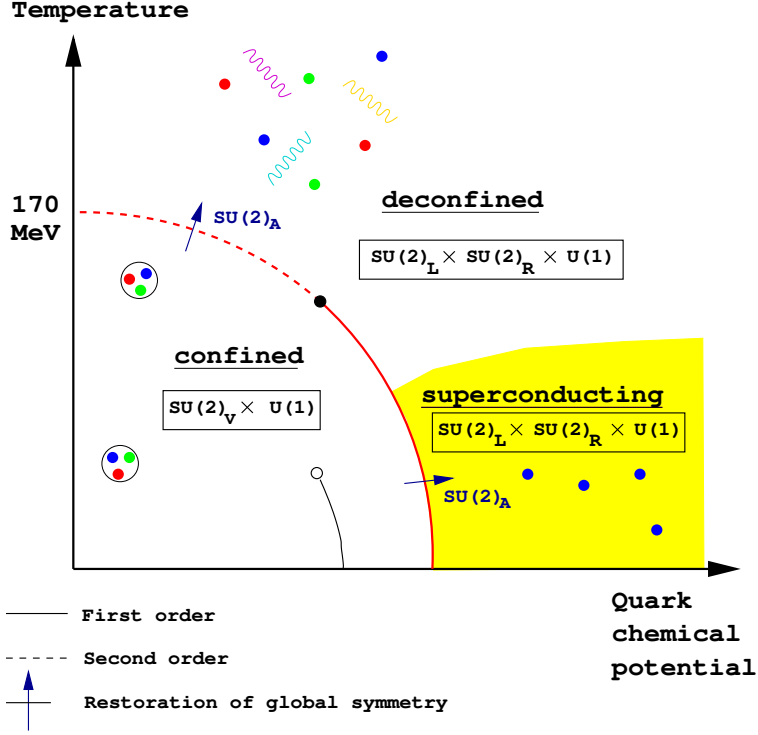,width=4in}
} 
\caption{Two massless flavor phase diagram}
\label{fig:2flav}
\end{figure}

In the real world there are two light quark flavors, the up 
($u$) and down ($d$), with
masses $\lesssim 10~\MeV$, and a medium-weight flavor, the strange 
($s$) quark, with mass $\sim 100~\MeV$.
A first approximation is to ignore the strange, and set $m_{u,d}=0$.

The gap equation for this scenario has been solved
using various interactions: pointlike four-fermion
interactions based on the instanton vertex 
\cite{Barrois,ARW2,RappETC,BergesRajagopal},
a full instanton vertex including all the form factors
\cite{CarterDiakonov}, and a weakly coupled gluon propagator
\cite{SW-pert,PR-Tc,PR-pert,HMSW,rockefeller,Hsu2}.
All agree that the quarks prefer to
pair in the color ${\bf \bar 3}$ flavor singlet
channel
\beq
\label{2flav:vev}
\mbox{2SC~phase:}\quad
\De^{\al\be}_{ij} =
\< q^\al_i  q^\be_j \>^{\phantom\ad}_{1PI} \propto C \ga_5\ep_{ij}\ep^{\al\be 3}
\eeq
(color indices $\al,\be$ run from 1 to 3, flavor indices 
$i,j$ run from 1 to 2).
The four-fermion interaction calculations also agree on the
magnitude of $\De$: around $100~\MeV$. This is found to be
roughly independent of the cutoff, although the chemical potential
at which it is attained is not.
Such calculations are based on
calibrating the coupling to give a chiral condensate of around
$400~\MeV$ at zero density, 
and turning $\mu$ up to look for the maximum gap.

As with any spontaneous symmetry breaking, one of the degenerate
ground states is arbitrarily selected.
In this case, quarks of the first two colors
(red and green) participate in pairing, while the third color
(blue) does not.
The ground state is invariant under an $SU(2)$
subgroup of the color rotations that mixes
red and green, but the blue quarks are singled out as different.
The pattern of symmetry breaking is therefore (with gauge symmetries
in square brackets)
\beq
\label{2flav:syms}
\ba{rl}
& [SU(3)_{\rm color}]\times [U(1)_Q]
 \times SU(2)_L \times SU(2)_R \\
\longrightarrow & 
 [SU(2)_{\rm color}]\times [U(1)_{\Qt}]
 \times SU(2)_L \times SU(2)_R \\
\ea
\eeq
The expected phase diagram in the $\mu$-$T$ plane is shown in
Figure~\ref{fig:2flav}.
The features of this pattern of condensation are
\begin{itemize}
\item The color gauge group is broken down to $SU(2)$, so five of the gluons
will become massive, with masses of order the gap (since the coupling
is of order~1). The remaining three gluons are associated with an
unbroken $SU(2)$ red-green gauge symmetry, whose confinement 
distance scale rises exponentially with density~\cite{SU2unbroken}.
\item
The red and green quark modes acquire a gap $\De$, which is the mass of the
physical excitations around the Fermi surface (quasiquarks).
There is no gap for the blue quarks in this ansatz, and it is an
interesting question whether they find some other channel in which to pair.
The available attractive channels appear to be weak
so the gap will be much smaller, perhaps
in the \keV\ range~\cite{ARW2,1SC}. It has even been suggested
that 'tHooft anomaly matching may prevent any condensation
\cite{Sannino, SanHsu}.
\item Electromagnetism is broken, but this does not mean that the 2SC phase
is an electromagnetic as well as a color superconductor.
Just as in the standard model the Higgs VEV leaves unbroken
a linear combination $Q$ of the weak $W_3$ and hypercharge $Y$ bosons,
so here a linear combination $\Qt$ of the eighth gluon $T_8$ and the
electric charge $Q$ is left unbroken.
This plays the role of a ``rotated'' electromagnetism.
We will discuss some of its physical effects in a later section.
\item No global symmetries are broken
(although additional condensates that break chirality have been 
suggested~\cite{Berges})
so the 2SC phase has the same
symmetries as the quark-gluon plasma (QGP), 
so there need not be any phase transition between them.
Again, this is in close analogy to the physics of the standard model, where
the Higgs VEV breaks no global symmetries: the phase transition line between
the unbroken and broken phases ends at some critical Higgs mass, and
the two regimes are analytically connected.
The reader may wonder why one cannot construct
an order parameter to distinguish
the 2SC phase using the fact that the
quark pair condensate blatantly breaks baryon number,
which is a global symmetry. However, in the two flavor case
baryon number is a linear combination of
electric charge and isospin, $B = 2Q - 2I_3$, so baryon number is already 
included in the symmetry groups of \Eqn{2flav:syms}. Just as an admixture
of gluon and photon survives unbroken as a rotated electromagnetism,
so an admixture of $B$ and $T_8$ survives unbroken as a rotated baryon number.
\end{itemize}

\section{THREE MASSLESS QUARK FLAVORS}
\label{sec:3flavor}

\begin{figure}[htb]
\centerline{
 \psfig{figure=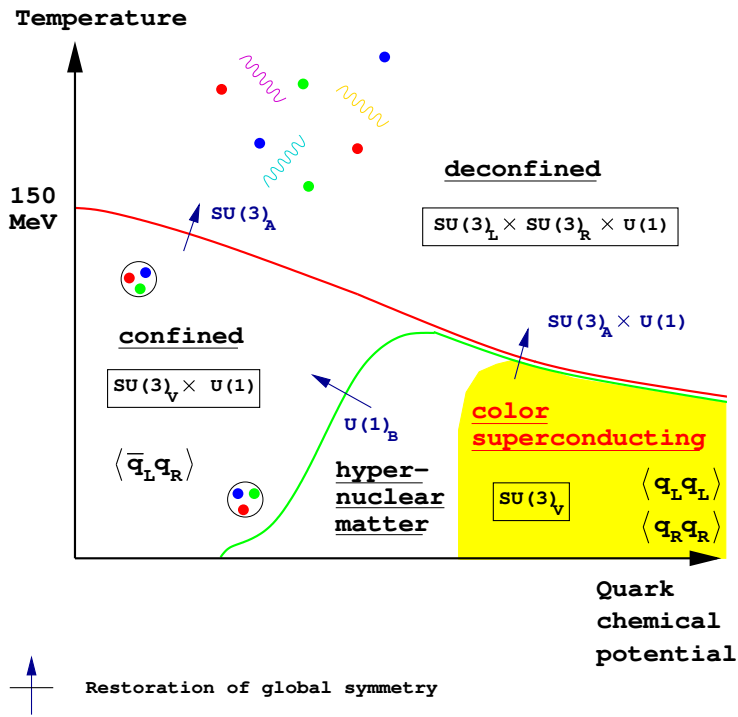,width=4in}} 
\caption{Three massless flavor phase diagram}
\label{fig:3flav}
\end{figure}

In QCD with three flavors of massless quarks 
the Cooper pairs {\it cannot}
be flavor singlets, and both color and flavor symmetries are
necessarily broken~\cite{ARW3} (see also~\cite{SS} for zero density).
The gap equation has been solved
for pointlike 4-fermion interactions with the index structure
of single gluon exchange~\cite{ARW3,SW-cont,HsuCFL} as well as
a weakly coupled gluon propagator
\cite{SchaeferPatterns,ShovWij}.
They agree that the attractive channel 
exhibits a pattern
called color-flavor locking (CFL),

\beq
\ba{rcl}
\mbox{CFL~phase:}\quad
\De^{\al\be}_{ij} = \< q^\al_i q^\be_j \>^{\phantom\ad}_{1PI}
&\propto& 
 C \ga_5 [\ep^{\al\be X}\ep_{ijX} 
 + \ka (\de^\al_i\de^\be_j + \de^\al_j\de^\be_i)] \\
&\propto& C \ga_5 [ (\ka+1)\de^\al_i\de^\be_j + (\ka-1) \de^\al_j\de^\be_i]
\ea
\label{3flav:vev}
\eeq
(color indices $\al,\be$ and flavor indices 
$i,j$ all run from 1 to 3).
The first line shows the connection between this and the 2SC pairing
pattern.
When $\ka=0$ the pairing is in the $(\bar{\bf 3}_A,\bar{\bf 3}_A)$
channel of color and flavor, which corresponds to three 
orthogonal copies of the 2SC pairing: the red and green
$u$ and $d$ pair as in 2SC, in addition the red and blue
$u$ and $s$ pair, and finally the green and blue $d$ and $s$ pair.
The term multiplied by $\ka$ corresponds to pairing in the
$({\bf 6}_S,{\bf 6}_S)$. It turns out that this additional condensate,
although not highly favored energetically (the color ${\bf 6}_S$ is
not attractive in single gluon exchange, instanton vertex, or 
strong coupling) breaks no additional symmetries and so
$\ka$ is in general small but not zero~\cite{ARW3,PisarskiCFL}.
A weak-coupling calculation~\cite{SchaeferPatterns} shows that
$\ka$ is suppresed by one power of the coupling,
$\ka = g \sqrt{2}\log(2)/(36\pi)$.

The second line of \Eqn{3flav:vev} exhibits the color-flavor locking
property of this ground state. The Kronecker deltas dot
color indices with flavor indices, so that the VEV is not
invariant under color rotations, nor under flavor rotations,
but only under simultaneous, equal and opposite, color and flavor
rotations. Since color is only a vector symmetry, this
VEV is only invariant under vector flavor rotations, and
breaks chiral symmetry.

The pattern of symmetry breaking is therefore
\beq
[SU(3)_{\rm color}]
 \times \underbrace{SU(3)_L \times SU(3)_R}_{\dsp\supset [U(1)_Q]}
 \times U(1)_B 
\longrightarrow \underbrace{SU(3)_{C+L+R}}_{\dsp\supset [U(1)_{\Qt}]}
 \times \Z_2 
\eeq
The expected phase diagram in the $\mu$-$T$ plane is shown in
Figure~\ref{fig:3flav}.
The features of this pattern of condensation are
\begin{itemize}
\item The color gauge group is completely broken. All eight gluons
become massive. This ensures that there are no infrared divergences
associated with gluon propagators.
\item
All the quark modes are gapped. The nine quasiquarks 
(three colors times three flavors) fall into an ${\bf 8} \oplus {\bf 1}$
of the unbroken global $SU(3)$, so there are two
gap parameters. The singlet has a larger gap than the octet.
\item Electromagnetism is no longer a separate
symmetry, but corresponds to gauging one of the flavor generators.
A rotated electromagnetism (``$\Qt$'')
survives unbroken. Just as in the 2SC case it is a combination
of the original photon and one of the gluons, although the
relative coefficients are different.
\item Two global symmetries are broken,
the chiral symmetry and baryon number, so there are two 
gauge-invariant order parameters
that distinguish the CFL phase from the QGP,
and corresponding Goldstone bosons which are long-wavelength
disturbances of the order parameter. 
The order parameter for the chiral symmetry is
$\<\psibar_L \ga_\mu\la^A\psi_L \psibar_R \ga_\mu\la^A\psi_R\>$
where $\la^A$ are the flavor generators
\cite{SchaeferPatterns}
(which only gets a vacuum expectation value beyond 
the mean field approximation).
The chiral Goldstone bosons
form a pseudoscalar octet, 
like the zero-density $SU(3)_{\rm flavor}$ pion octet.
The breaking of the baryon number symmetry has order parameter
$\<udsuds\> = \<\La\La\>$, and a singlet scalar Goldstone boson
which makes the CFL phase a superfluid.

If a quark mass were introduced then it would explicitly break
the chiral symmetry and give a mass
to the chiral Goldstone octet, but the CFL phase would still be 
a superfluid, distinguished by its baryon number breaking.

\item
Quark-hadron continuity. It is striking that the symmetries of the
3-flavor CFL phase are the same as those one might expect for 3-flavor
hypernuclear matter~\cite{SW-cont}. In hypernuclear matter one
would expect the hyperons to pair in an $SU(3)_{\rm flavor}$
singlet ($\<\La\La\>, \<\Si\Si\>, \< N \Xi \>$), breaking baryon number but
leaving flavor and electromagnetism unbroken. Chiral symmetry would
be broken by the chiral condensate. This means that one might be able to
follow the spectrum continuously from hypernuclear matter to
the CFL phase of quark matter---there need be no phase transition.
The pions would evolve into the pseudoscalar octet mentioned above.
The vector mesons would evolve into the massive gauge bosons.
This will be discussed in more
detail below for the 2+1 flavor case.
\end{itemize}
 
We can now draw a hypothetical phase diagram for 3-flavor QCD
(Figure~\ref{fig:3flav}).
Comparing with the 2-flavor case, we see that the 2SC quark-paired
phase is easy to distinguish from nuclear matter, since it has
restored chiral symmetry, but hard to distinguish from the QGP.
The CFL phase is easy to distinguish from the QGP, but hard
to distinguish from hypernuclear matter.

We conclude that dense quark matter has rather different global
symmetries for $m_s=0$ than for $m_s=\infty$. Since the real world
has a strange quark of middling mass, it is very interesting to
see what happens as one interpolates between these extremes.

\begin{table}[t]
\caption{Symmetries of phases of QCD.}
\vspace{.5pc}
\begin{center}
\bt{cccc}
phase & electromagnetism & chiral symmetry & baryon number \\
\hline
QGP & $Q$ & unbroken & $B$ \\
\hline
\bt{c} 2 flavor\\[-0.5ex] nuclear matter\et 
  & broken & broken & broken \\
\hline
\bt{c} 2 flavor quark\\[-0.5ex]  pairing (2SC) \et
  & $\Qt=Q-\frac{1}{2\sqrt{3}}\Y$ & unbroken   & $\tilde B = \Qt + I_3$ \\
\hline
\bt{c} 3 flavor\\[-0.5ex]  nuclear matter \et 
  & $Q$ & broken & broken \\
\hline
\bt{c} 3 flavor quark\\[-0.5ex]  pairing (CFL) \et
  & $\Qt=Q+\frac{1}{\sqrt{3}}\Y$   & broken  & broken \\
\et
\end{center}
\end{table}

\begin{figure}[thb]
\begin{center}
 \psfig{figure=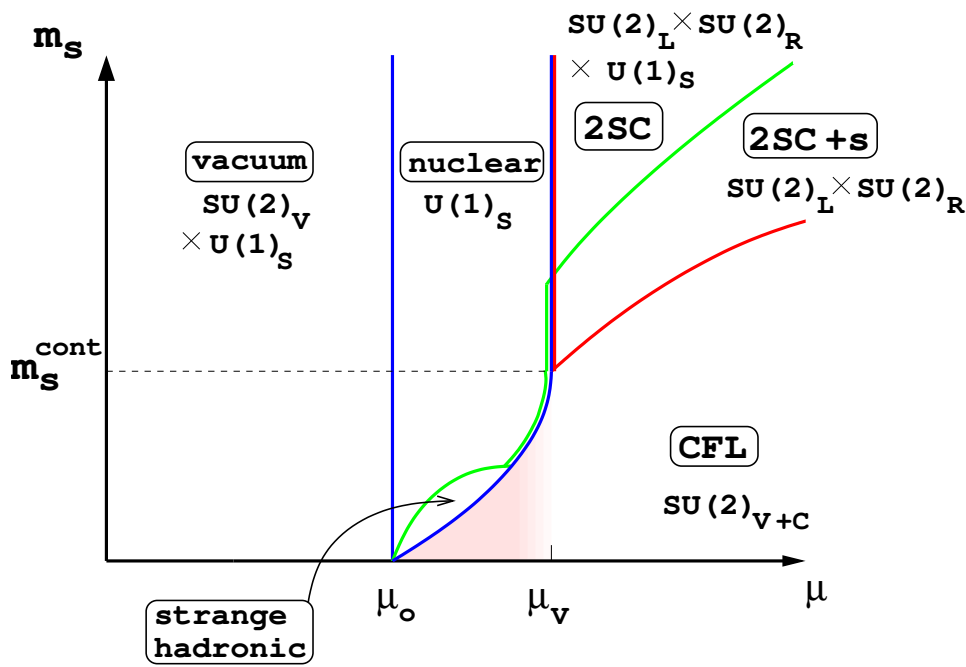,width=5in}
\end{center}
\caption{Conjectured phase diagram for 2+1 flavor QCD at $T=0$.
The global symmetries of each phase are labelled.
The red line marks chiral symmetry breaking, the blue line
isospin breaking, and the green line strangeness breaking.
The regions of the phase diagram labelled 
2SC, 2SC+s and CFL denote color superconducting quark matter phases.
The pink shading marks the region of quark-hadron continuity.
A detailed explanation is given in the text.
}
\label{fig:qcd2+1flav}
\end{figure}

\section{TWO MASSLESS + ONE MASSIVE QUARK FLAVORS}
\label{sec:2+1flavor}

A nonzero strange quark mass explicitly breaks the 
flavor $SU(3)_L\times SU(3)_R$ symmetry down to
$SU(2)_L\times SU(2)_R$. 
If the strange quark is heavy enough then it will decouple,
and 2SC pairing will occur.
For a sufficiently small strange 
quark mass we expect a reduced form of color-flavor 
locking in which an $SU(2)$ subgroup of $SU(3)_{\rm color}$
locks to isospin, causing chiral symmetry breaking and
leaving a global $SU(2)_{{\rm color}+V}$ group unbroken.

As $m_s$ is increased from zero to infinity, there has to be some 
critical value at which the strange quark decouples,
color and flavor rotations are unlocked, and 
the full $SU(2)_L \times SU(2)_R$ symmetry is restored.
It can be argued on general grounds (see below) that
a simple unlocking phase transition must be first order, although
there are strong indications that there is
a crystalline intermediate
phase (see Section \ref{sec:unlock}).

An analysis of the unlocking transition, using a NJL model with
interaction based on single-gluon exchange~\cite{ABR2+1,SW-cont}
confirms this expectation. Although the quantitative results from
NJL models can only be regarded as rough approximations, it is interesting
that the calculations indicate
that for realistic values of the strange quark mass 
chiral symmetry breaking may be present for
densities all the way down to those characteristic of baryonic matter.
This raises the possibility that
quark matter and baryonic matter may be continuously 
connected in nature, as Sch\"afer and Wilczek 
have conjectured for QCD with three massless quarks~\cite{SW-cont}.
The gaps due to pairing at the quark
Fermi surfaces map onto gaps
due to pairing at the baryon Fermi surfaces in 
superfluid baryonic matter consisting of nucleons, 
$\Lambda$'s, $\Sigma$'s, 
and $\Xi$'s (see below).

Based on the NJL calculations, the zero-temperature
phase diagram as a function of chemical potential and 
strange quark mass
has been conjectured~\cite{ABR2+1} to be as shown in 
Figure~\ref{fig:qcd2+1flav}.
Electromagnetism was ignored in this calculation, 
and it was assumed that wherever a baryon Fermi surface is
present, baryons always pair at zero temperature.
To simplify our analysis, we assume that baryons always
pair in channels which preserve rotational invariance,
breaking internal symmetries such as isospin if necessary. 
So the real
phase diagram may well be even more complicated.


We characterize the phases using the $SU(2)_L\times SU(2)_R$ 
flavor rotations of the light quarks, and the 
$U(1)_S$ rotations of the strange quarks. The
$U(1)_B$ symmetry associated
with baryon number is a combination of $U(1)_S$, a $U(1)$ subgroup of
isospin, and the gauged $U(1)_{\rm EM}$ of electromagnetism.
Therefore, in our analysis of the global symmetries, 
once we have analyzed isospin and strangeness,
considering baryon number adds nothing new.

\subsection{Description of the phase diagram}
To explain Figure~\ref{fig:qcd2+1flav}, we follow the phases
that occur from low to high density, first for large $m_s$, then
small $m_s$.

\subsubsection{Heavy strange quark}
For $\mu=0$ the density is
zero; isospin and strangeness are unbroken; Lorentz symmetry is
unbroken; chiral symmetry is broken.
Above a first order transition~\cite{Halasz} at an onset chemical
potential $\mu_{\rm o}\sim 300~\MeV$, one finds nuclear matter.
Lorentz symmetry is broken, leaving only rotational symmetry manifest.  
Chiral symmetry is broken, though perhaps with a reduced
chiral condensate. We expect an instability of the nucleon Fermi
surfaces to lead to Cooper pairing, and assume that, as is observed in
nuclei, the pairing is $pp$ and $nn$, breaking isospin
(and perhaps also rotational invariance).
Since there
are no strange baryons present, $U(1)_S$ is unbroken.
When $\mu$ is
increased above $\mu_{\rm V}$, we find the ``2SC'' phase of
color-superconducting matter consisting of up and down quarks only,
paired in Lorentz singlet isosinglet channels. The full
flavor symmetry $SU(2)_L\times SU(2)_R$ is restored.
The phase transition at $\mu_V$ is first order
according to NJL models with low cutoff
\cite{BergesRajagopal,BJW2,PisarskiRischke1OPT,CarterDiakonov}
and random matrix models~\cite{StephanovRandMat}
as the chiral condensate competes with the superconducting
condensate.

When $\mu$ exceeds the
constituent strange quark mass $M_s(\mu)$, a strange quark Fermi
surface forms, with a Fermi momentum far below that for the light
quarks.  The strange quarks pair with each other, in a color-spin
locked phase~\cite{Schaefer1Flavor} that we call ``2SC+s''.
Strangeness is now broken, but the  $ss$
condensate is expected to be small~\cite{Schaefer1Flavor}.

Finally, when the chemical potential is high enough that the Fermi
momenta for the strange and light quarks become comparable, we cross
into the color-flavor locked (CFL) phase.
There is an unbroken global symmetry constructed by locking the
$SU(2)_V$ isospin rotations and an $SU(2)$ subgroup of color. Chiral
symmetry is once again broken.

\subsubsection{Light strange quark}
Below $\mu_{\rm o}$, we have the vacuum, as before.
At $\mu_{\rm o}$, one enters the nuclear matter phase, with the 
familiar $nn$ and $pp$ pairing at the neutron and proton 
Fermi surfaces breaking isospin.  

At a somewhat larger chemical potential, strangeness is broken, first
perhaps by kaon condensation 
\cite{KaplanNelson,BrownRho,SchaeferKaon}
or by the appearance and Cooper pairing
of strange baryons, $\La$ and $\Si$, and then $\Xi$, which
pair with themselves in spin singlets.
This phase is labelled ``strange hadronic'' in
Figure~\ref{fig:qcd2+1flav}.  The global symmetries $SU(2)_L\times
SU(2)_R$ and $U(1)_S$ are all broken.   

We can imagine two possibilities for what happens next as $\mu$ increases
further, and we enter the pink region of the figure.
(1)~Deconfinement: the baryonic Fermi surface is replaced by
$u,d,s$ quark Fermi
surfaces, which are unstable against pairing, and
we enter the CFL phase, described above. Isospin is locked to color and
$SU(2)_{{\rm color}+V}$ is restored, but chiral symmetry remains broken.
(2)~No~deconfinement: the Fermi momenta of all of the octet
baryons are now similar enough that pairing between baryons with
differing strangeness becomes possible.  At this point,
isospin is restored: the baryons pair in rotationally
invariant isosinglets
($p\Xi^-$, $n\Xi^0$, $\Si^+ \Si^-$, $\Si^0\Si^0$, $\La \La$).
The interesting point is that scenario (1) and scenario (2) are 
indistinguishable.
Both look like the ``CFL'' phase of the figure:
$U(1)_S$ and chirality are broken, and there is an
unbroken vector $SU(2)$. This is
the ``continuity of quark and hadron matter''
described by Sch\"afer and Wilczek~\cite{SW-cont}.
We conclude that for low enough strange quark mass, $\ms<\ms^{\rm cont}$, there
may be a region where sufficiently dense baryonic matter has the same
symmetries as quark
matter, and there need not be any 
phase transition between them. 

Color-flavor 
locking will always occur 
for sufficiently large chemical potential, for 
any nonzero, finite $m_s$.  This follows from
Son's model-independent analysis valid at 
very high densities.\cite{Son}
As a consequence of color-flavor locking,
chiral symmetry is spontaneously broken even at asymptotically
high densities, in sharp contrast to the well established
restoration of chiral symmetry at high temperature.

Finally, it is interesting to ask what we expect at non-zero
temperature. There has been no comprehensive NJL study of
this, but one can make the reasonable guess that
quark pairing with a gap $\De$ at $T=0$
will disappear in a phase transition at $T_c\approx 0.6 \De$.
This is the BCS result, which is also found to hold
for quark pairing~\cite{PR-Tc,Hong-pert}.

\begin{figure}[htb]
\hbox{
  \psfig{figure=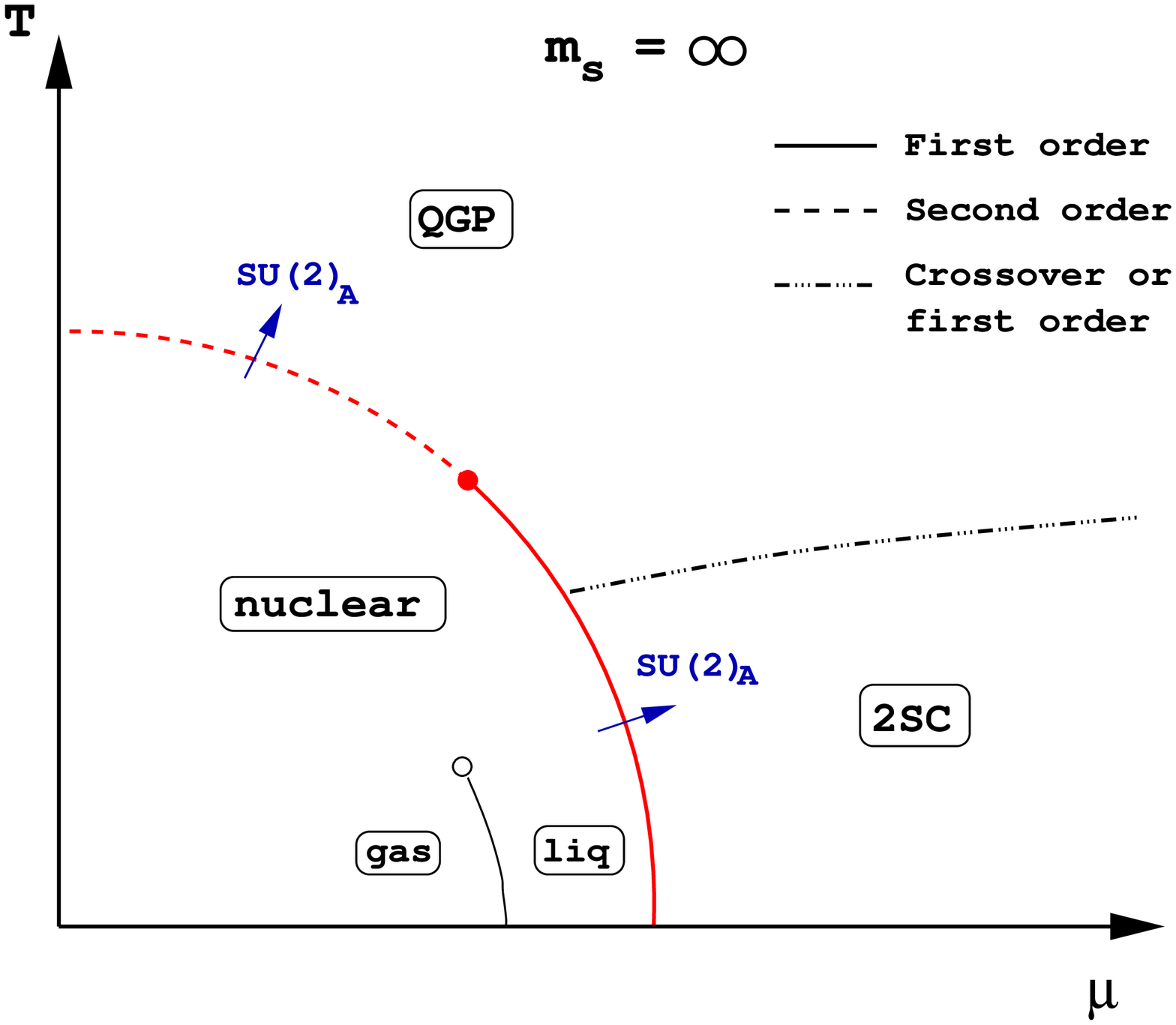,width=2.5in} 
  \psfig{figure=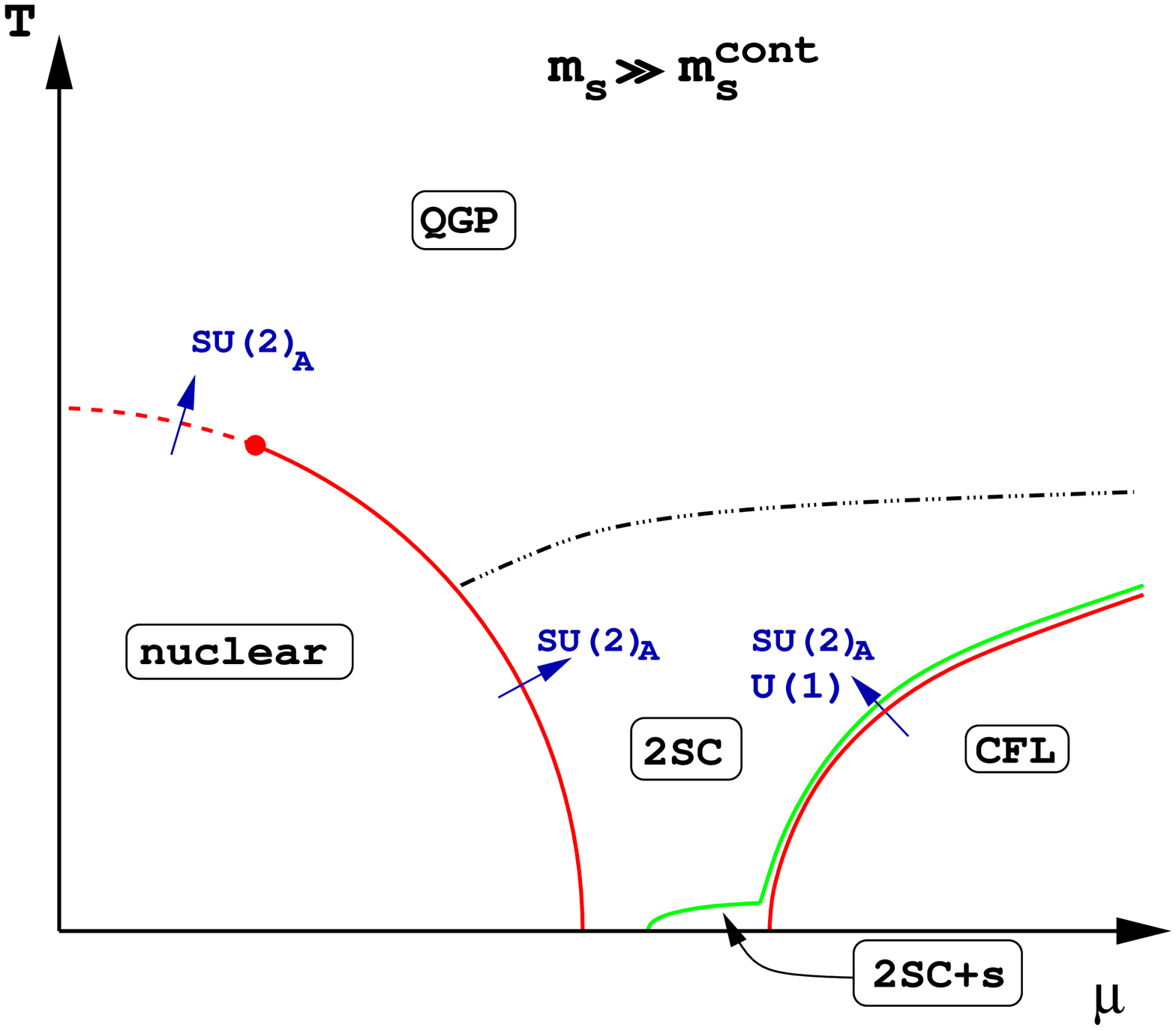,width=2.5in}
}
\hbox{ 
 \psfig{figure=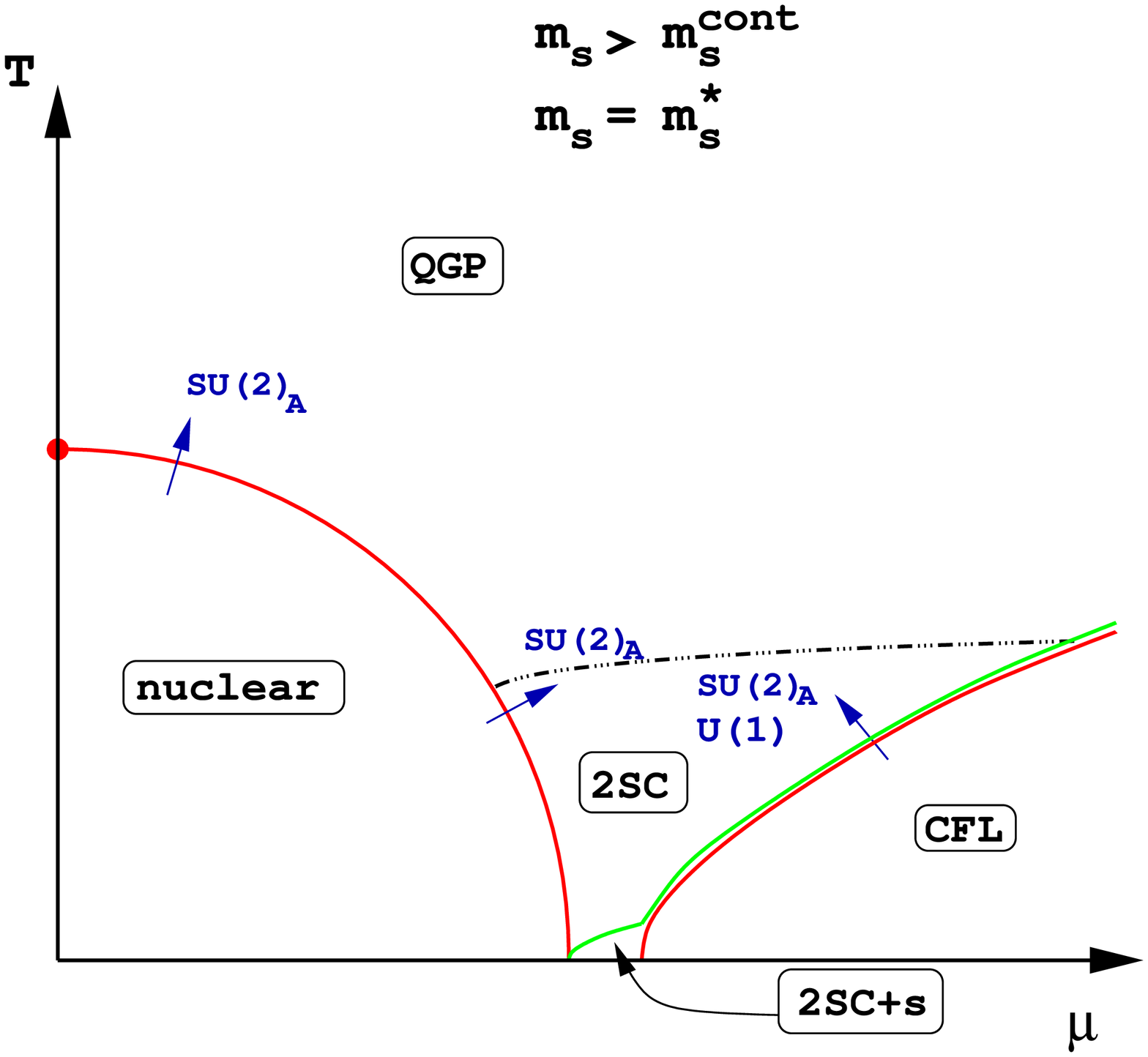,width=2.5in}
 \psfig{figure=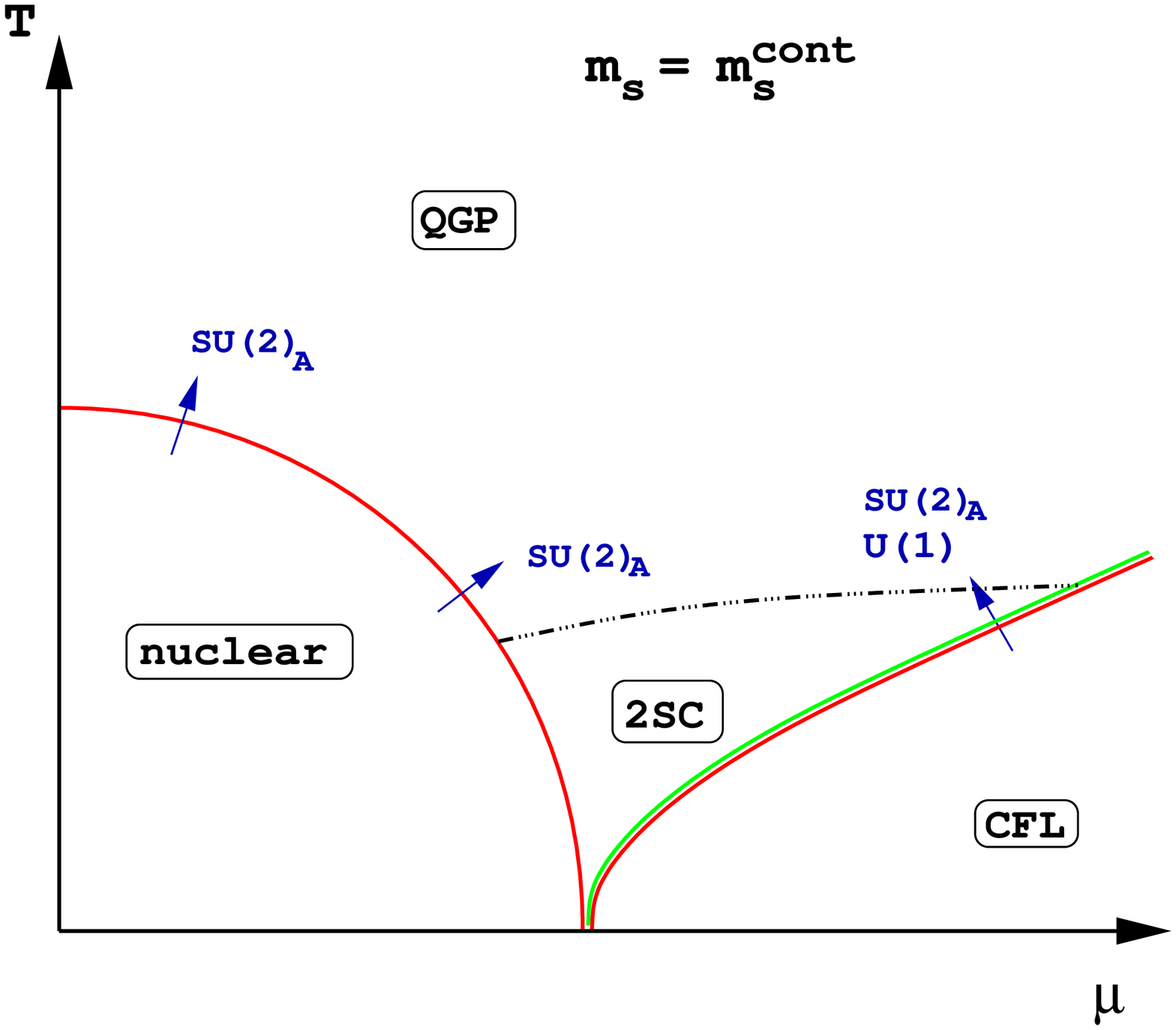,width=2.5in}
} 
\hbox{
 \psfig{figure=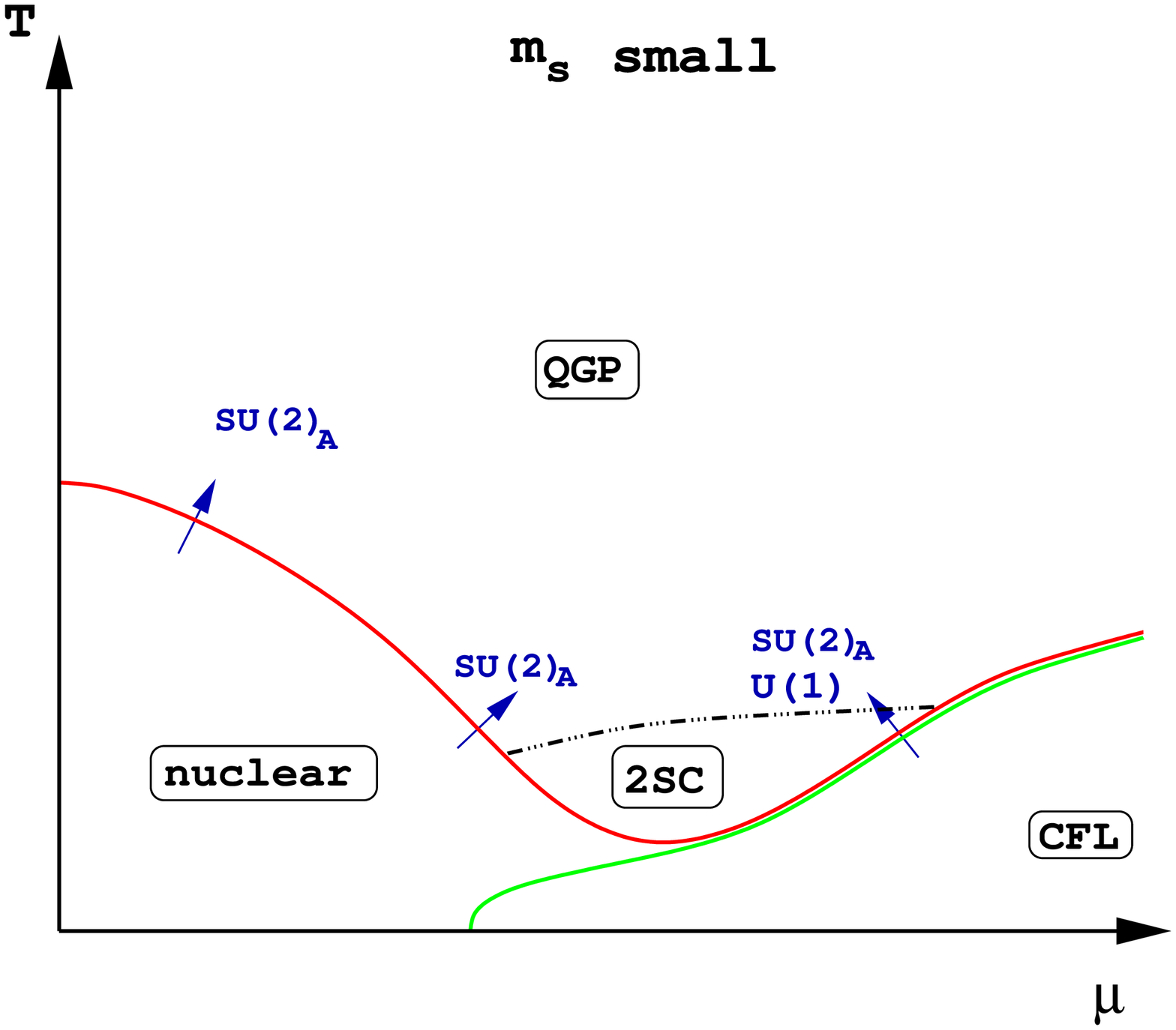,width=2.5in} 
 \psfig{figure=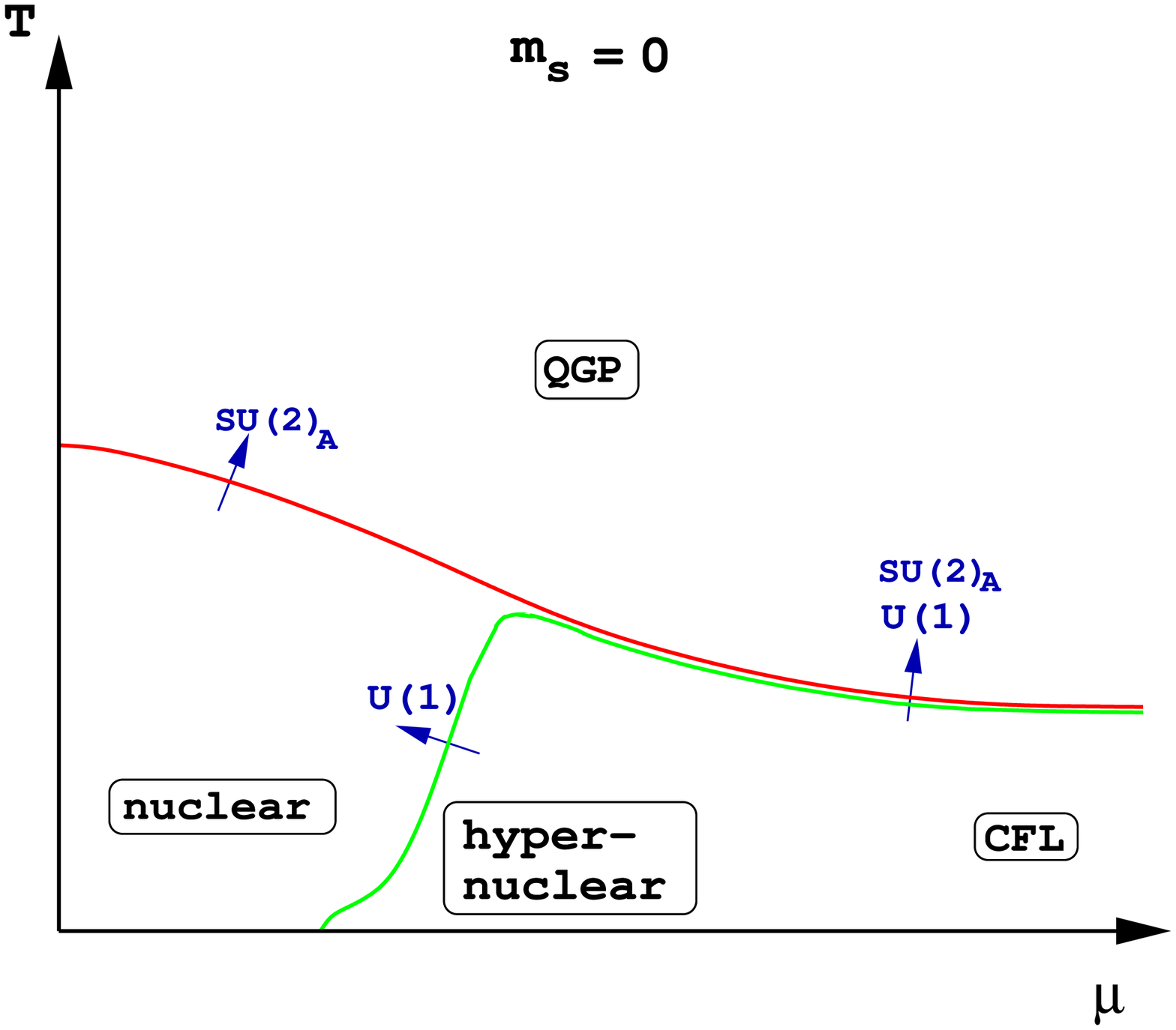,width=2.5in}
} 
\caption{ 
The phases of QCD in the chemical potential and temperature
plane, for various strange quark masses, interpolating from
the two flavor to the three flavor theory. The up and down quark
masses and electromagnetism are neglected. Lines of breaking of
$SU(2)$ chiral symmetry and $U(1)$ strangeness are shown.
The QGP and 2SC phases have the same global symmetries, so the dot-dash
line separating them represents the possibility of
first order transitions that become crossovers
at critical points.
} 
\label{fig:sections}
\end{figure}

\begin{figure}[thb]
\begin{center}
 \psfig{figure=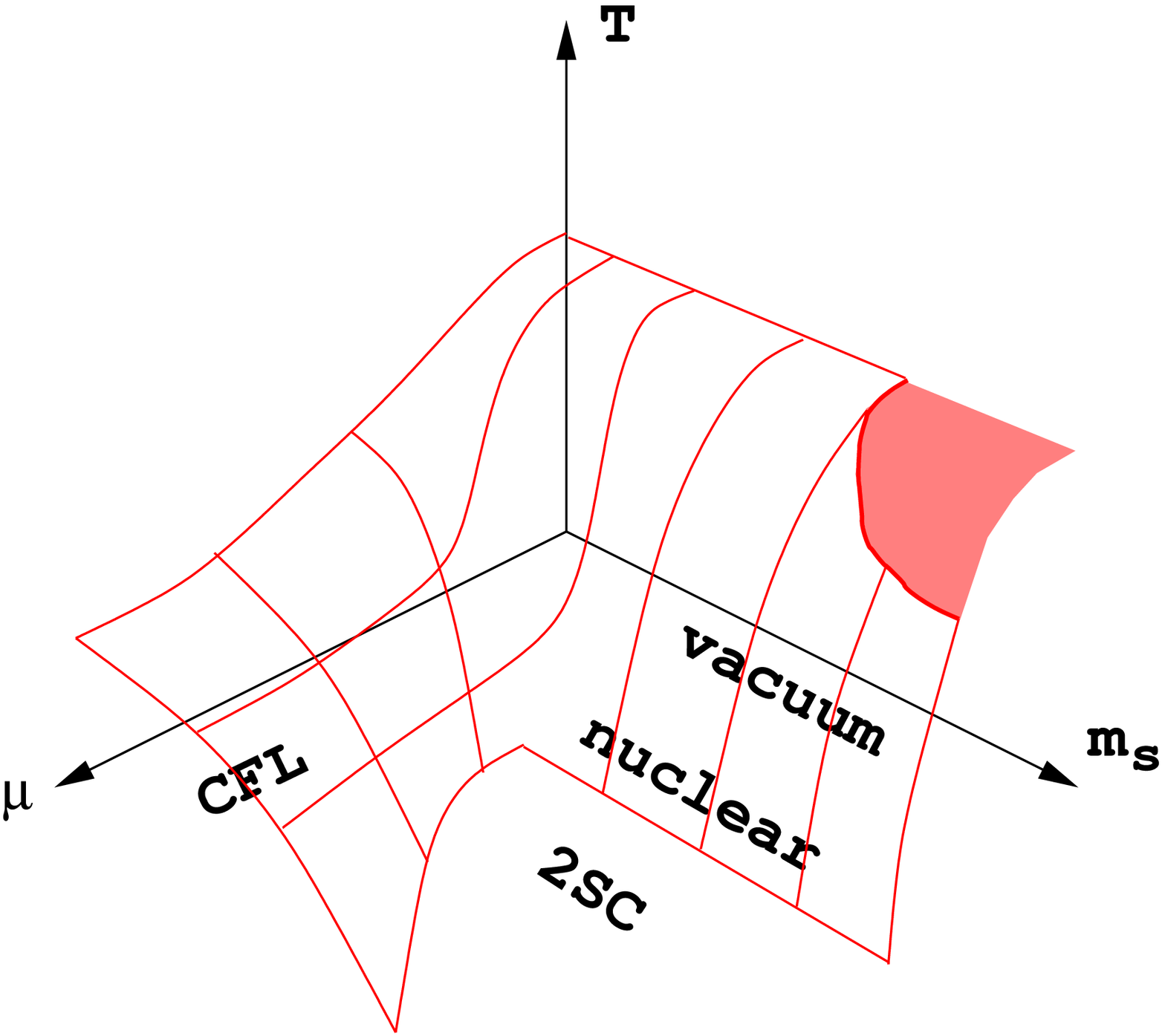,width=5in}
\end{center}
\caption{
The chiral phase transition surface as a function of chemical potential
$\mu$, strange quark mass $\ms$, and temperature $T$.
The diagrams of Figure~\ref{fig:sections} are a sequence of 
$\mu$-$T$ sections through this space.
The shaded region at high $\ms$ and $T$
is the second-order part of the critical surface,
which is bounded by a tricritical line. Everywhere else the phase
transition is conjectured to be first-order.
}
\label{fig:chiral3d}
\end{figure}

Assuming the zero-temperature
phase structure given in Figure~\ref{fig:qcd2+1flav}, we can
guess that the non-zero temperature $\mu$-$T$ phase diagram
for strange quark masses varying from infinity to zero
will be as shown in the diagrams of Figure~\ref{fig:sections}.
These are assembled into a single three-dimensional
diagram in Figure \ref{fig:chiral3d}, where for clarity
only the chiral phase transition surface is shown:
the thick red line is tricritical, and the red shaded
region that it bounds is second-order.

The main features of the phase diagram are as follows.
\begin{itemize}
\item The second-order chiral phase transition 
(dashed line) 
that is present at low density and high temperature
shrinks as the strange quark becomes lighter, until at $\ms=\ms^*$
the tricritical point arrives at $T=0$. At lower masses,
there is no second-order line.
\item The strangeness-breaking line (green) and the high-density
chiral symmetry breaking line (red)
do not exactly coincide because at low enough temperature there
is a window of densities where strange quarks are present, but
their Fermi momentum is too low to allow them to pair with
the light quarks. This is the 2SC+s phase, where the strange quarks
pair with themselves,
breaking strangeness/baryon number, in a color-spin locked
phase  whose gap and critical temperature
are very small~\cite{Schaefer1Flavor}.
\item At arbitrarily high densities, where the QCD gauge coupling is small,
quark matter is always in the CFL phase with broken chiral symmetry. This is
true independent of whether the ``transition'' to quark matter is
continuous,  or whether, as for larger $\ms$, there are two first
order transitions, from nuclear matter to the 2SC phase, and
then to the CFL phase.
\item
Color-flavor locking survives for $M_s^2 \lesssim 2\sqrt{2} \mu\De$
(see below). Since the CFL state is $\Qt$-neutral, 
there are no electrons present in this phase~\cite{CFLneutral},
so introducing electromagnetism makes no difference to it.
\end{itemize}
Additional features, beyond those required by symmetry considerations
alone,  have been suggested by Pisarski~\cite{Pisarski}, by
analogy with scalar-gauge field theories.

\subsection{Quark-hadron continuity}
\label{sec:cont}
The pink region in Figure~\ref{fig:qcd2+1flav} is characterized
by a definite global symmetry, $SU(2)\times [U(1)]$, but
this can either be a hadronic (hyperonic) phase with unbroken isospin and
electromagnetism, or a color-flavor locked quark matter phase with 
an isospin+color symmetry and a rotated electromagnetism
that allows a linear combination of the photon and a gluon
to remain massless. In other words, in this
regime there is no symmetry difference
between hyperonic matter and quark matter.
This raises the exciting possibility
\cite{SW-cont} that properties of sufficiently dense 
hadronic matter could be found by extrapolation from the quark matter
regime where weak-coupling methods can be used.

\begin{table}[htb]
\caption{Quark hadron continuity: mapping between states in 
high density hadronic and quark matter.}
\begin{tabular}{cc@{$\quad\Leftrightarrow\quad$}c}
\hline
Particle type & Hyperonic matter & CFL quark matter \\
\hline
Fermions: & 8 Baryons & 9 Quarks \\
Chiral (pseudo)Goldstone: & 8 pion/kaons & 8 pseudoscalars \\
Baryon number (pseudo)Goldstone: & 1 & 1 \\
Vector Mesons: & 9 & 8 massive gluons \\
\hline
\end{tabular}
\label{tab:states}
\end{table}

The most straightforward application of this idea is to relate the
quark/gluon description of the spectrum to the hadron description of
the spectrum in the CFL phase.\cite{SW-cont} The conjectured
mapping is given in Table~\ref{tab:states}.
Gluons in the CFL phase map to the octet of vector
bosons; the Goldstone bosons
associated with chiral symmetry breaking in the CFL phase map to the
pions; and the quarks map onto baryons. Pairing occurs at the Fermi
surfaces, and we therefore expect the gap parameters in the various
quark channels to map to the gap parameters due to baryon pairing.

\begin{table}[htb]
\newlength{\wid}\settowidth{\wid}{XXX}
\def\st{\rule[-1.5ex]{0em}{4ex}}
\caption{Mapping of fermionic states between high density quark 
and hadronic matter.}
\begin{tabular}{lcc|ccc} 
\hline
\st Quark & $SU(2)_{{\rm color}+V}$ & $\Qt$ & 
   Hadron & $SU(2)_{V}$   & $Q$ \\
\hline
\multirow{2}{\wid}{$\left(\ba{c} bu\\[1ex] bd \ea\right)$} & 
\multirow{2}{2em}{\bf 2} &
$+1$ &
\multirow{2}{4em}{$\left(\ba{c} p\\[1ex] n \ea\right)$} & 
\multirow{2}{2em}{\bf 2} &
$+1$
\\
& & 0 & & & 0 \st \\
\multirow{2}{\wid}{$\left(\ba{c} gs\\[1ex] rs \ea\right)$} & 
\multirow{2}{2em}{\bf 2} &
0 &
\multirow{2}{4em}{$\left(\ba{c} \Xi^0\! \\[1ex] \Xi^-\!\! \ea\right)$} & 
\multirow{2}{2em}{\bf 2} &
0 \st \\
& & $-1$ & & & $-1$ \st \\
\hline
\multirow{3}{\wid}{$\left(\ba{c} ru-gd\\[1ex] gu\\[1ex] rd \ea\right)$} & 
\multirow{3}{2em}{\bf 3} &
0 & 
\multirow{3}{4em}{$\left(\ba{c} \Si^0 \\[1ex] \Si^+ \\[1ex] \Si^- \ea\right)$} & 
\multirow{3}{2em}{\bf 3} &
0
\\
& & $+1$ & & & $+1$ \st \\
& & $-1$ & & & $-1$ \st \\
\hline
$ru+gd+\xi_- bs$\hspace{-1em} & \hspace{-2em} {\bf 1} & 0 & 
  $\La$ \hspace{-1em} & \hspace{-2em} {\bf 1} & 0 
\\
\hline
$ru+gd-\xi_+ bs$ & \hspace{-2em}{\bf 1} & 0 & 
  --- &  \st \\
\hline
\end{tabular}
\label{tab:qm}
\end{table}

In Table~\ref{tab:qm}
we show how this works for the fermionic states in 2+1 flavor QCD.
There are nine states in the quark matter phase. We show how they
transform under the unbroken ``isospin'' of $SU(2)_{{\rm color}+V}$ and their
charges under the unbroken ``rotated electromagnetism'' generated
by $\Qt$, as described in Section 4. 
Table~\ref{tab:qm} also shows the baryon octet,
and their transformation properties under the symmetries
of isospin and electromagnetism that are unbroken in sufficiently
dense hadronic matter. Clearly there is a correspondence between
the two sets of particles.\footnote{The
one exception is the final isosinglet. 
In the $\mu\to\infty$ limit, where the
full 3-flavor symmetry is restored, it becomes an $SU(3)$ singlet,
so it is not expected to map to any member of the baryon octet.
The gap $\De_+$ in this channel is twice as large as the others.
}
As $\mu$ increases, 
the spectrum described in Table 2 may evolve continuously 
even as the language used to describe it changes from baryons, 
$SU(2)_{V}$ and $Q$ to quarks, $SU(2)_{{\rm color}+V}$ and $\Qt$.

If the spectrum changes continuously, then in particular so must the
gaps.  As discussed above,
the quarks pair into rotationally invariant,
$\Qt$-neutral, $SU(2)_{{\rm color}+V}$ singlets.  The two doublets of
Table \ref{tab:qm} pair with each other, the triplet pairs with itself.
Finally, the two singlets pair with themselves.

\section{COLOR-FLAVOR UNLOCKING AND THE CRYSTALLINE
COLOR SUPERCONDUCTING PHASE}
\label{sec:unlock}

A prominent feature of the zero temperature phase diagram
Figure~\ref{fig:qcd2+1flav} is the ``unlocking'' phase transition
between two-flavor pairing (2SC) and three-flavor pairing (CFL).
At this phase transition, the Fermi momentum of free
strange quarks is sufficiently different from that
of the light quarks to disrupt pairing between them.

Such transitions are expected to be a generic feature of
quark matter in nature.
In the absence of interactions, the requirements of weak equilibrium
and charge neutrality cause all three flavors of quark to have 
different Fermi momenta.
In the extreme case where all three flavors had very
different chemical potentials, each flavor would have to self-pair
\cite{Iwa3flav,Schaefer1Flavor}, but in the phenomenologically
interesting density range we expect a rich and complex phase structure for
cold dense matter as a function of quark masses and density.
The CFL $\leftrightarrow$ 2SC transition of Figure~\ref{fig:qcd2+1flav}
is one example. 
Assuming that no other intermediate
phases are involved, we now give a model independent argument that the 
unlocking phase transition between the CFL and 2SC phases
in Figure~\ref{fig:qcd2+1flav} must be first order.
However, there is good reason to expect an intermediate state---the
crystalline color superconducting state, and we go on to discuss it
in some detail.
Note that another crystalline phase, the ``chiral crsytal''
has also been proposed~\cite{RappCrystal}, although it
is not yet clear whether there is any window of density where
it is favored.

\subsection{The (un)locking transition}

\begin{figure}[t]
\begin{center}
\psfig{figure=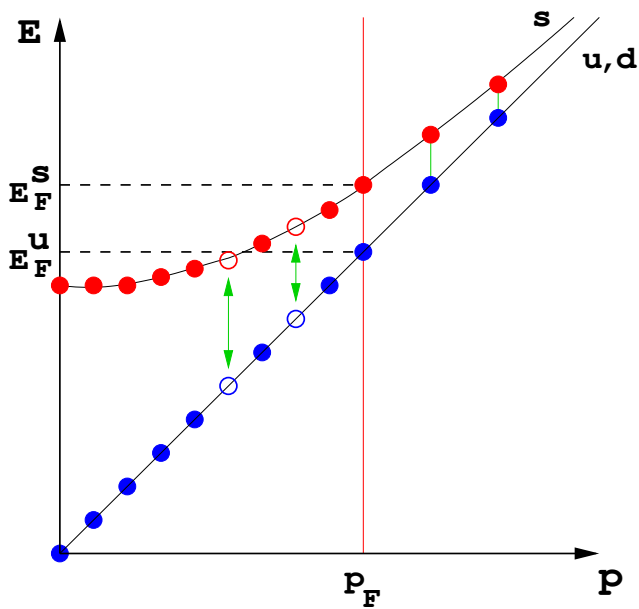,width=3.5in}
\end{center}
\caption{
How the strange quark mass interferes with
a $u$-$s$ condensate.  The strange
quark (upper curve) and light quark (straight line) dispersion
relations are shown, with their Fermi seas filled up to a common Fermi
momentum $p_F$.  The horizontal axis is the magnitude of the
spatial momentum; $s$-wave pairing occurs between particles (or holes)
with the same $p$ and opposite $\vec p$.
The energy gained by pairing stops the $s$ quarks from decaying to $u$
quarks (see text).
}
\label{fig:fermi}
\end{figure}

Figure~\ref{fig:fermi} shows part of the CFL pairing pattern: the
quark states of the different flavors are filled up
to a common Fermi momentum $p_F\approx \mu$, intermediate between
the free light quark and free strange quark Fermi momenta.
In the absence of interactions, this state would be unstable:
weak interactions would turn strange quarks into light quarks,
and there would be separate strange and light Fermi momenta,
each filled up to the Fermi energy $\mu$.
However, pairing stabilizes it.
For the paired state to be stable, it must be that the
free energy gained from turning a strange quark into a light quark
is less than the energy lost by breaking the Cooper pairs
for the modes involved~\cite{CFLneutral}.
\begin{equation}
\ba{c}
\dsp \sqrt{\mu^2+M_s(\mu)^2} - \sqrt{ \mu^2+M_u(\mu)^2} 
 \approx \frac{M_s(\mu)^2-M_u(\mu)^2}{2\mu} 
\lesssim 2\Delta_{us}\ , \\[1ex]
\dsp \mbox{i.e.},~~\frac{M_s(\mu)^2}{4\mu} \lesssim \Delta_{us}
\ea
\label{criterion}
\end{equation}
Here $M_s(\mu)$ and $M_u(\mu)$ are the constituent quark masses
in the CFL phase, and $M_u(\mu)\ll M_s(\mu)$.
An additional factor of $1/\sqrt{2}$ on the RHS of \eqn{criterion}
can be obtained requiring the paired state to have lower free energy
\cite{Clogston,OurLOFF,CFLneutral}.
Equation (\ref{criterion})
implies that arbitrarily small values of $\Delta_{us}$ are
impossible, which means that the phase transition must
be first-order: the gap cannot go continuously to zero.
Such behavior has been found in calculations
for unlocking phase transitions of this kind 
in electron superconductors~\cite{Clogston} and
nuclear superfluids~\cite{Sedrakian} as well as QCD
superconductors~\cite{ABR2+1,SW2,Bedaque}.

\subsection{The crystalline color superconducting phase}

There is good reason to think that, in the region where the strange
quark is just on the edge of decoupling from the light quarks, another
form of pairing can occur.  This is the ``LOFF'' state, first explored
by Larkin and Ovchinnikov~\cite{LO} and Fulde and Ferrell~\cite{FF} in
the context of electron superconductivity in the presence of magnetic
impurities.  They found that near the unpairing transition, it is
favorable to form a state in which the Cooper pairs have nonzero
momentum. This is favored because it gives rise to a region of phase
space where each of the two quarks in a pair can be close to its Fermi
surface, and such pairs can be created at low cost in free energy.
Condensates of this sort spontaneously break translational and
rotational invariance, leading to gaps which vary periodically in a
crystalline pattern.  
The possible consequences for compact stars will be discussed in
section \ref{sec:compact}.


In Ref.~\cite{OurLOFF}, the LOFF phase in QCD has been studied
using a toy model in which the
quarks interact via a
four-fermion interaction
with the quantum numbers of single gluon exchange.
The model only considers pairing between $u$ and $d$ quarks, with
$\mu_d=\bar\mu+\delta\mu$ and $\mu_u=\bar\mu-\delta\mu$.
For the rest of this section we will discuss properties of the model,
but it is important to remember that in reality
we expect a LOFF state wherever the difference between the Fermi momenta
of any two quark flavors is near an unpairing transition,
for example the unlocking phase transition between the 2SC and
CFL phases.  

\subsubsection{The nature of LOFF pairing}

\begin{figure}
\begin{center}
\psfig{figure=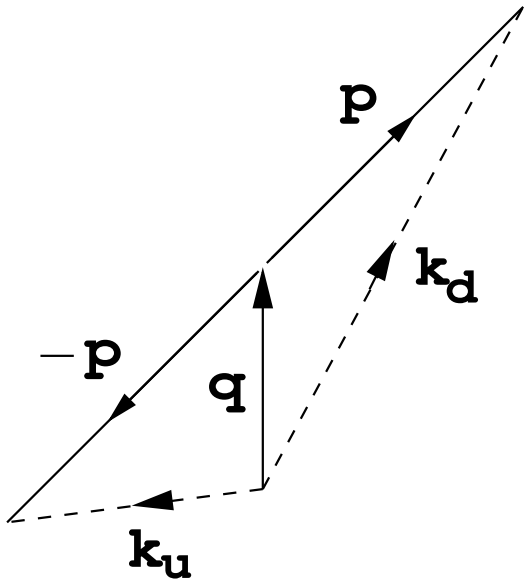,width=3.0in}
\end{center}
\caption{
The momenta $\vk_u$ and $\vk_d$
of the two members of a LOFF-state Cooper pair.  We choose
the vector $\vq$, common to all Cooper pairs,
to coincide with the $z$-axis.  
}
\label{fig:angles}
\end{figure}

Whereas the BCS state requires pairing
between fermions with equal and opposite momenta, for some
values of $\dm$
it may be more favorable to form a condensate of Cooper pairs with
{\it nonzero} total momentum.  By pairing quarks with momenta which
are not equal and opposite, some Cooper pairs are allowed to have both
the up and down quarks on their respective Fermi surfaces even
when $\dm\neq 0$.  LOFF found that within a range of $\dm$ 
a condensate of Cooper pairs with momenta $\vk_u =\vq+\vp$ and
$\vk_d=\vq-\vp$ (see Figure~\ref{fig:angles}) is favored over either the BCS
condensate or the normal state. Here, our notation is such that $\vp$
specifies a particular Cooper pair, while $\vq$ is a fixed vector,
the same for all pairs,
which characterizes a given LOFF state.  The magnitude $|\vq|$ is
determined by minimizing the free energy; the direction of $\vq$ is
chosen spontaneously.  The resulting LOFF state breaks translational
and rotational invariance.  In position space, it describes a
condensate which varies as a plane wave with wave vector $2\vq$.

\subsubsection{Results from a simplified model}
\begin{figure}[t]
\parbox{2.3in}{
\psfig{figure=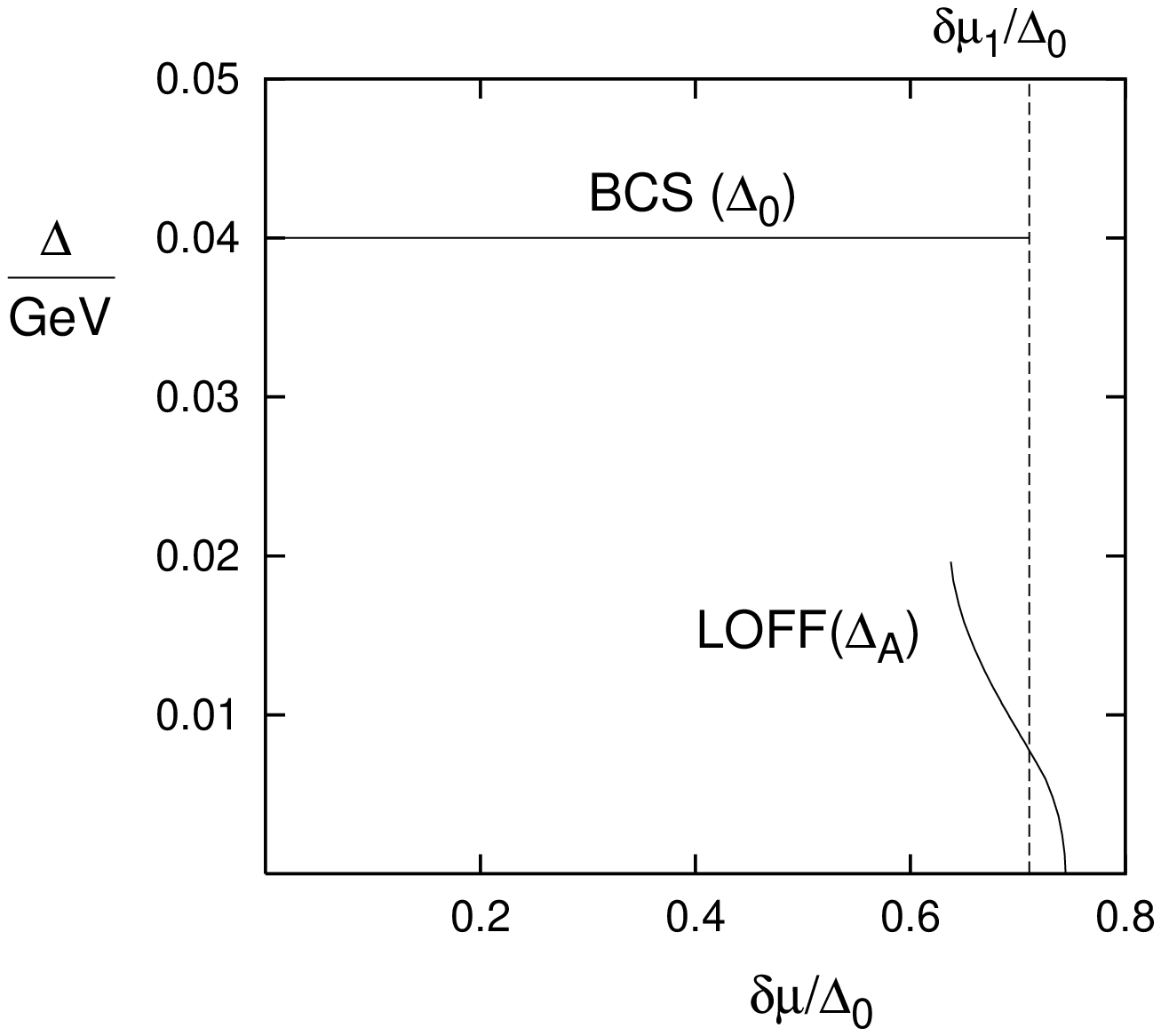,height=2.3in,angle=-90}
}
\parbox{2.3in}{ 
\psfig{figure=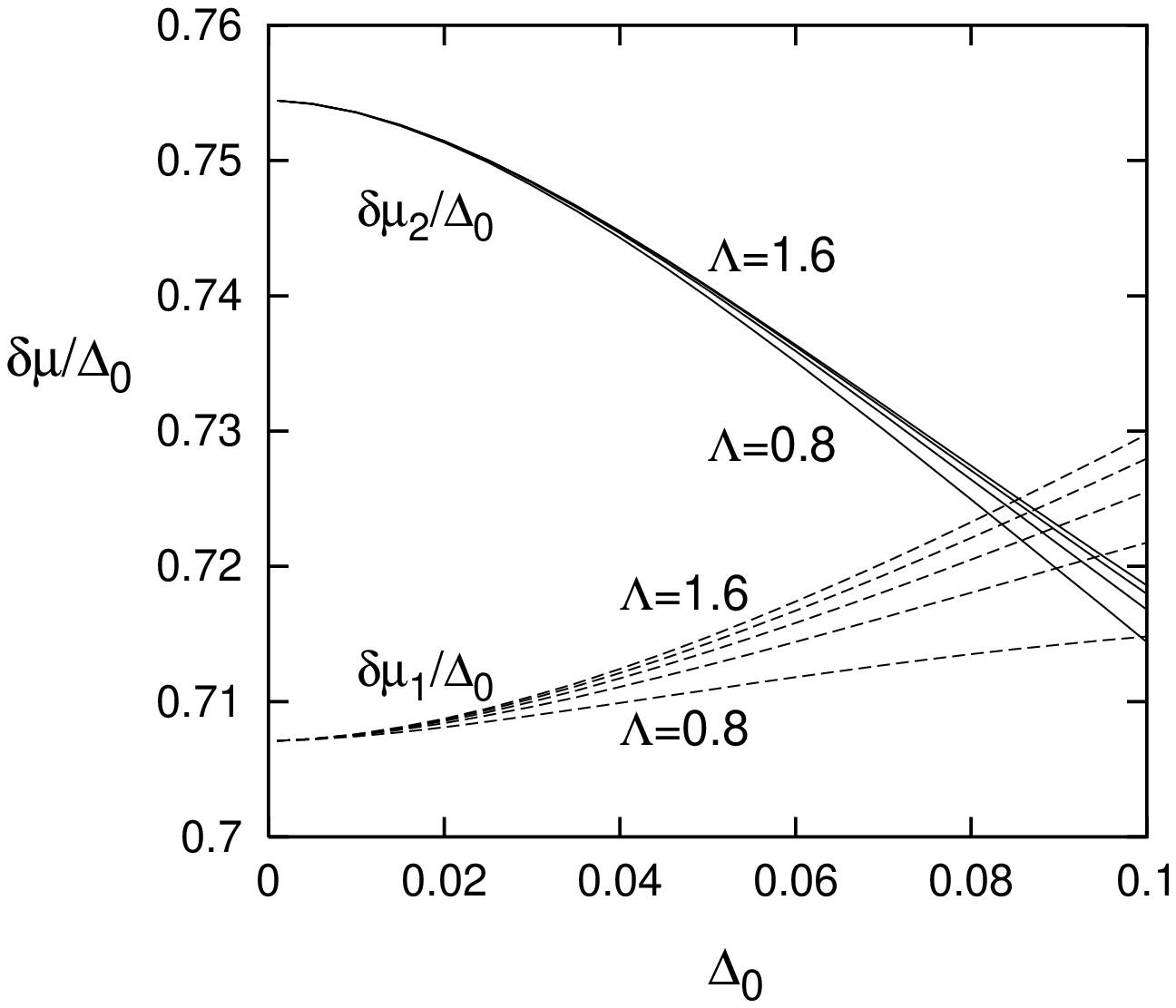,height=2.3in,angle=-90}
}
\caption{(a) LOFF and BCS gap parameters as a function of $\delta\mu$,
with coupling chosen so that $\Delta_0=40~\MeV$, and
$\La=1~\GeV$.    The vertical dashed line marks 
$\delta\mu=\delta\mu_1$, above
which the LOFF state has lower free energy than BCS.
(b) The interval of $\delta\mu$ within which the LOFF state occurs
as a function of the coupling, 
parametrized by the BCS gap $\Delta_0$ in GeV. 
Below the solid line, there is a LOFF state. Below the dashed line,
the BCS state is favored. The different lines of each type correspond to
different cutoffs on the four-fermion interaction:
$\Lambda=0.8~\GeV$ to $1.6~\GeV$. 
$\delta\mu_1/\Delta_0$ and $\delta\mu_2/\Delta_0$ show little
cutoff-dependence, and the cutoff dependence disappears completely
as $\Delta_0,\delta\mu\rightarrow 0$.
}
\vspace{-0.1in}
\label{fig:fish}
\end{figure}

In the LOFF state, each Cooper pair carries 
momentum $2{\bf q}$, so the condensate and gap parameter 
vary in space with wavelength $\pi/|{\bf q}|$. In the range of $\dm$
where the LOFF state is favored, $|{\bf q}|\approx 1.2 \delta\mu$.
In Ref.~\cite{OurLOFF}, we simplify
the calculation of the gap parameter
by assuming that the condensate varies in space
like a plane wave, leaving the determination of the crystal
structure of the QCD LOFF phase to future work. 
We make an ansatz for the LOFF wave function,
and by variation obtain a gap equation which allows
us to solve for the gap parameter $\Delta_A$, the free energy and
the values of the diquark condensates which characterize
the LOFF state at a given $\delta\mu$ and $|{\bf q}|$. 
We then vary the momentum $|{\bf q}|$
of the ansatz, to find the preferred (lowest
free energy) LOFF state at a given $\delta\mu$, and compare
the free energy of the LOFF state to that of the BCS state with
which it competes. We show results for one choice
of parameters in Figure~\ref{fig:fish}(a).

In Figure~\ref{fig:fish} the
average quark chemical potential $\bar\mu$ has been 
set to 0.4~GeV, corresponding to
a baryon density of about 4 to 5 times that in nuclear matter.
A crude estimate~\cite{OurLOFF} suggests that 
in quark matter at this density, $\delta\mu\sim 15-30$~MeV
depending on the value of the density-dependent effective
strange quark mass.  

We find that the LOFF state is favored
for values of $\delta\mu$ which satisfy 
$\delta\mu_1 < \delta\mu < \delta\mu_2$ as shown in Figure~\ref{fig:fish}(b),
with $\delta\mu_1/\Delta_0=0.707$ and $\delta\mu_2/\Delta_0=0.754$ in the 
weak coupling limit $\Delta_0\ll \mu$.  ($\Delta_0$
is the 2SC gap for $\delta\mu<\delta\mu_1$, and one can
use it to parametrize the strength of the four fermion interaction $G$:
small $\Delta_0$ corresponds to a small $G$.)
At weak coupling, the LOFF gap parameter decreases from $0.23 \Delta_0$
at $\delta\mu=\delta\mu_1$ (where there is a first order BCS-LOFF
phase transition)
to zero at $\delta\mu=\delta\mu_2$ (where there is a second order
LOFF-normal transition).  
Except for very close to $\delta\mu_2$, the critical
temperature above which the LOFF state melts will be much
higher than typical neutron star temperatures.
At stronger coupling the LOFF gap parameter decreases relative
to $\Delta_0$ and 
the window of $\delta\mu/\Delta_0$ within which the LOFF state
is favored shrinks, as seen in Figure~\ref{fig:fish}(b).  

\subsubsection{General properties of the LOFF phase}
The LOFF state is characterized by a gap parameter $\Delta_A$ and a 
diquark condensate, but not by an energy gap in the dispersion
relation. Because the condensate breaks rotational invariance,
the quasiparticle dispersion relations vary
with the direction of the momentum, yielding gaps that vary from zero
up to a maximum of $\Delta_A$. 

Near the second-order critical point $\delta\mu_2$, we can describe the
phase transition with a Ginzburg-Landau effective potential.
The order parameter for the LOFF-to-normal phase transition is
\begin{equation}\label{lofforderparam}
\Phi({\bf r}) = -\frac{1}{2} \langle \epsilon_{ab} \epsilon_{\alpha\beta3} 
\psi^{a\alpha}({\bf r}) C \gamma_5 \psi^{b\beta}({\bf r}) \rangle 
\end{equation}
so that in the normal phase $\Phi({\bf r}) = 0$, while in the LOFF phase
$\Phi({\bf r}) = \Gamma_A e^{i 2 {\bf q} \cdot {\bf r}}$.  (The gap parameter
is related to the order parameter by $\Delta_A=G\Gamma_A$.)
Expressing the order
parameter in terms of its Fourier modes $\tilde\Phi({\bf k})$, we write
the LOFF free energy (relative to the normal state) as
\begin{equation}\label{ginzland}
F(\{\tilde\Phi({\bf k})\}) = \sum_{{\bf k}} \left( C_2(k^2) | 
\tilde\Phi({\bf k}) |^2 
+ C_4(k^2) | \tilde\Phi({\bf k}) |^4 + {\mathcal O}(|\tilde\Phi|^6) \right).
\end{equation}
For $\delta\mu > \delta\mu_2$, 
all of the $C_2(k^2)$ are positive and the normal
state is stable.  Just below the critical point, all of the modes
$\tilde\Phi({\bf k})$ are stable except those on the sphere $|{\bf k}| =
2q_2$, where $q_2$ is the value of $|{\bf q}|$ at $\delta\mu_2$ 
(so that $q_2\simeq 1.2 \delta\mu_2 \simeq 0.9 \Delta_0$ 
at weak coupling).  In general,
many modes on this sphere can become nonzero, giving a
condensate with a complex crystal structure.  We consider the simplest
case of a plane wave condensate where only the one mode
$\tilde\Phi({\bf k} = 2{\bf q}_2) = \Gamma_A$ is nonvanishing.  Dropping all
other modes, we have
\begin{equation}\label{ginzland2}
F(\Gamma_A) = a(\delta\mu - \delta\mu_2) (\Gamma_A)^2 + b (\Gamma_A)^4 
\end{equation}
where $a$ and $b$ are positive constants.  Finding the minimum-energy
solution for $\delta\mu < \delta\mu_2$, we obtain simple power-law relations
for the condensate and the free energy:
\begin{equation}\label{powerlaws}
\Gamma_A(\delta\mu) = K_{\Gamma} (\delta\mu_2 - \delta\mu)^{1/2}, 
\hspace{0.3in} F(\delta\mu) = - K_F ( \delta\mu_2- \delta\mu)^2.
\end{equation}
These expressions agree well with the numerical results obtained
by solving the gap equation~\cite{OurLOFF}.  
The Ginzburg-Landau
method does not specify the proportionality factors $K_\Gamma$ and
$K_F$, but analytical expressions for these coefficients can be
obtained in the weak coupling limit by explicitly solving the gap
equation~\cite{Takada1,OurLOFF}.

Notice that because $(\delta\mu_2-\delta\mu_1)/\delta\mu_2$ is small, the
power-law relations (\ref{powerlaws}) are a good model of the system
throughout the entire LOFF 
interval $\delta\mu_1 < \delta\mu < \delta\mu_2$ where the
LOFF phase is favored over the BCS phase.  The Ginzburg-Landau
expression (\ref{ginzland2}) gives the free energy of the LOFF phase
near $\delta\mu_2$, but it cannot be used to determine the location
$\delta\mu_1$ of the first-order phase transition where the LOFF window
terminates, which requires a comparison of
LOFF and BCS free energies.

\section{COMPACT STAR PHENOMENOLOGY}
\label{sec:compact}

Having described the interesting phenomena that we believe occur
in cold quark matter, we now ask ourselves where 
in nature such phenomena might
occur, and how we might see evidence of them.

The only place in the universe where we expect sufficiently
high densities and low temperatures is compact stars,
also known as ``neutron stars'', since it is often
assumed that they are made primarily of neutrons 
(for a recent review, see~\cite{HP-nstar}).
A compact star is produced in a supernova. As the outer layers
of the star are blown off into space, the core collapses
into a very dense object. Typical compact stars have masses
close to $1.4 \Msolar$, and are believed to have radii of order 10 km.
The density ranges from around nuclear density near the surface
to higher values further in, although uncertainty about the equation
of state leaves us unsure of the value in the core.

During the supernova, the core collapses, and its gravitational energy
heats it to temperatures of order $10^{11}~\Kelvin$ (tens of \MeV),
but it cools rapidly by neutrino emission.
Within a few minutes its internal temperature $T$ drops to
$10^9$~K ($100~\keV$), and reaches $10^7$~K ($1~\keV$) after a century.
Neutrino cooling continues to dominate for the first
million years of the life of the star.
The effective temperature $T_e$ of the X-ray emissions
is lower than the internal temperature: 
$T_e/10^6\,\Kelvin \approx \sqrt{T/10^8\,\Kelvin}$~\cite{Page}.

Color superconductivity gives mass to
excitations around the ground state: it opens
up a gap at the quark Fermi surface, and makes the gluons
massive. One would therefore expect its main consequences
to relate to transport properties, such as mean free paths,
conductivities and viscosities.
The influence of color superconductivity on the equation of state
is an $\O((\De/\mu)^2)$ (few percent) effect, which is not phenomenologically
interesting given the existing uncertainty in the equation
of state at the relevant densities.

\subsection{Cooling by neutrino emission}

As mentioned above, for its first million or so years,
a neutron star cools by neutrino emission. 
The temperature is obtained from X-ray spectra of isolated compact stars,
and is subject to many uncertainties, including emissions from
plasma around the star, and distortion of the spectrum by 
a possible hydrogen atmosphere.
The age, inferred from the spindown rate by assuming magnetic dipole
radiation from a constant dipole moment, may also have large
systematic errors. Even so, a consistent picture emerges
\cite{Schaab,Page} in which the youngest compact stars,
about a thousand years old, have surface temperatures 
around $2\times 10^6$~K ($200~\eV$),
falling to about $3\times 10^5$~K ($30~\eV$) after 
a million years.

The cooling rate is
determined by the heat capacity and emissivity, both
of which are dominated by quark modes whose energy
is within $T$ of the Fermi surface, and are therefore
sensitive to the kind of gaps generated by
color superconductivity~\cite{Schaab,Blaschke,Page}.

In the CFL phase, all quarks and gluons have gaps $\Delta\gg T$,
electrons are absent~\cite{CFLneutral}, and the
transport properties are dominated by the
only true Goldstone excitation, the superfluid mode
arising from the breaking of the exact baryon number symmetry.
The next lightest modes are the pseudo-Goldstone bosons
associated with chiral symmetry breaking, which will
only participate when the temperature is above their mass,
which is
of order tens of  \MeV~\cite{effth}.
This means that CFL quark matter has a much smaller
neutrino emissivity and heat capacity than nuclear matter,
and hence the cooling of a neutron star is likely
to be dominated by the nuclear mantle
rather than the CFL core~\cite{Page}.
A CFL core is therefore not detectable by cooling measurements.

We turn now to the 2SC quark matter phase, which occurs
if the strange quarks are too heavy to pair with the light flavors.
Up and down quarks of two of the colors (red and green, say)
pair strongly with a gap much bigger than the temperature.
This leaves the blue up and down, and the strange quarks (if present)
with much more weakly attractive channels in which to pair.
The strange quarks are believed to pair with each other in
a ``color-spin locked'' condensate, with a gap of order
hundreds of keV~\cite{Schaefer1Flavor} or less~\cite{OurLOFF}.
The blue up and down quarks form $J=1$ pairs,
breaking rotational invariance~\cite{ARW2}, with a gap
that was originally estimated to be in the \keV\ range,
but this estimate is not robust, and depends on details of the 
NJL model used~\cite{ARW2}.

This leads to potentially interesting phenomenology,
since the blue and/or strange quarks have small gaps,
so during the early life of the compact star they
may participate in the cooling dynamics as long as the temperature
is greater than their gap. Their effects would be dramatic,
allowing high rates of neutrino emission via direct URCA
processes such as $d\to u+e+\bar\nu$ and $u\to d+e^+ +\nu$,
and leading to rapid cooling of the core~\cite{Schaab,Page}.
The cooling would slow down suddenly when the temperature fell below the
gap. Such a behavior would be observable, and if no sign of it
is seen as our observations of neutron star temperatures improve
then we will have to conclude that either 2SC matter does
not occur, or the smallest gaps are larger than the
observed temperatures.

\subsection{The neutrino pulse at birth}

We have seen above that in the first seconds of a 
supernova, the inner regions
(``protoneutron star'') are heated to tens of \MeV\ by
the 
gain a vast amount of energy
from the gravitational collapse, and are consequently
hot (tens of \MeV). Over the next half-minute or so
much of the energy is radiated off as neutrinos,
whose detailed spectrum as a function of time
is determined by the neutrino diffusion properties
of the protoneutron star.
Neutrinos from supernova 1987A were detected in terrestrial
experiments, and the duration and mean energy of the
pulse was measured. We can hope that 
neutrinos from future supernovae
in our galaxy will be measured more precisely.
It is therefore useful to
study the effects of color superconductivity
on neutrino diffusion, in order to see if it leads
to any signature in the neutrino pulse.

Carter and Reddy~\cite{CarterReddy} have performed a
preliminary investigation of this question.
They restricted themselves to two flavors, and
studied the case where the core starts off as a hot
quark-gluon plasma. Within seconds,
thanks to neutrino emission, it cools into a superconducting phase,
and they assumed that this occured
via a second-order phase transition.
This leads to a striking two-stage signature:
(1) near the critical temperature $T_c$ the heat capacity
rises, and the cooling of the star consequently slows;
(2) below the critical temperature, the quark modes
are gapped, and the neutrino mean free path
is enhanced by $\exp(\De/T)$, reflecting Boltzmann suppression
of the population of quark quasiparticles.
As a result, the core may suddenly empty itself
of neutrinos, creating a final neutrino burst.
There may be further processing of this burst
on its way out of the supernova, but
the suggestion is that it may survive to yield a
noticeable signal in neutrino detectors on earth.
The suggestion, then, is that the flux of 
supernova neutrinos
detected on earth will not taper off, but show a final
burst followed by no flux. Before that, there may be
a plateau in the energy or flux of the neutrinos,
as the cooling slows near the critical temperature.

There are many issues that require further investigation.
It is not clear whether a second-order phase transition
is to be expected, since the up and down quark Fermi surfaces
will differ, and there may be a first-order unlocking
transition~\cite{Bedaque}. 
Also, it is necessary to take into account the strange quark,
and the processing of emitted neutrinos by the layers of
neutrino-opaque hadronic matter that surround the core during the
supernova explosion.

\subsection{r-mode instability}

\begin{figure}[htb]
\centerline{
 \psfig{figure=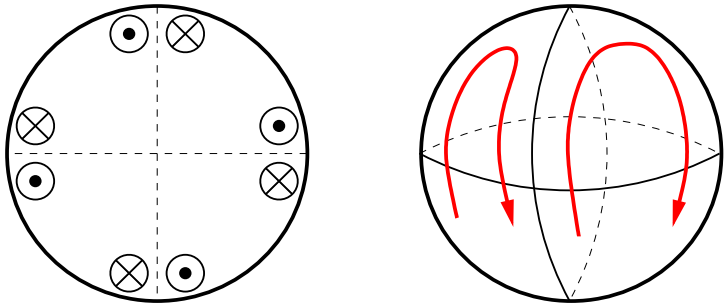,width=4in}} 
\caption{The quadrupole pattern of $r$-mode bulk flows}
\label{fig:rmode}
\end{figure}

The term ``$r$-mode'' (short for ``rotational mode'') refers to a bulk
flow in a rotating star that radiates away energy and angular momentum
in the form of gravitational waves
(Figure \ref{fig:rmode}).  If the rotation frequency $f$
of the star is above a critical value $f_*$, the system becomes
unstable to $r$-modes and will quickly spin down until its
frequency drops to $f_*$, at which point the $r$-modes are damped out.
The critical frequency depends on the sources of damping
that could suppress the flows. These include shear and bulk viscosities,
and also ``surface rubbing''---the friction at the interface between
the $r$-mode region and any rigid crust that cannot flow.
Since the viscosities are sensitive functions of temperature,
one calculates $f_*(T)$ as an upper limit on rotation
frequencies, and thereby maps out an excluded high-$f$ region
in the $T$-$f$ plane. Differently constituted compact stars
(neutrons, quarks with various gaps due to pairing)
have different excluded regions, and one can see
whether any of them are ruled out by the observation of
pulsars in nature
with rotation rates and temperatures in their excluded region.

Madsen~\cite{Madsen} has shown that gapless quark matter
and neutron stars are not ruled out. However,
color superconductivity
creates gaps in the quark excitation spectrum,
suppressing the viscosities by factors of
order $\exp(-\De/T)$, and
encouraging $r$-mode spindown.
He found that for a compact star made {\em entirely} of quark matter
in the CFL phase, even a quark gap as small as $\Delta=1$~MeV
reduces $f_*(T)$ dramatically to $\O(100~\Hz)$ for temperatures
below $10^9$~K ($100~\keV$). 
This means that millisecond pulsars, with frequencies up to 640~Hz,
cannot be CFL quark matter stars, making it questionable whether
any compact stars are made entirely of CFL quark matter.
Madsen found that 2SC quark matter stars were on the
edge of being ruled out, so he was not able to say anything
about them, either positive or negative.

Madsen included the additional damping from surface
rubbing between the quark matter and a normal matter crust.
Using the conventional picture, this is a very small effect,
since the crust is separated from the quark matter by an electrostatic
cushion of electrons, and so surface rubbing made no difference
to the result for pure CFL stars. Actually, since CFL matter is 
neutral~\cite{CFLneutral}, it contains no electrons,
so the cushioning mechanism may not be operative, and
it is not clear that there is any such crust.

There are caveats to Madsen's conclusions.
Firstly, the results are sensitive to the temperature
of the inner regions of the star, which has to be inferred
from the measured effective surface temperature using
models of the heat flow, and is therefore not accurately known.
However, this uncertainty is only important for unpaired or
2SC paired quark matter;
pure CFL stars are ruled out for $T_e<10^9~\Kelvin~(100~\keV)$,
which is already at the upper end of conceivable temperatures.
Secondly, as he points out, his calculations do not
rule out the generic picture of how quark
matter occurs in compact stars, namely as a
a quark matter core surrounded by a nuclear mantle.
In this case substantial friction is expected at the
core-mantle interface, and this may be enough
\cite{Bildsten,Madsen} to stabilize the
star irrespective of the viscosities of the quark matter.
Furthermore, quark matter may contain a shell of LOFF
crystal (see below), and the $r$-modes could be damped
at the edges of that region rather than at the crust.
We can hope that future work on hybrid stars will
clarify the situation.

\subsection{Magnetic field decay}

The behavior of magnetic fields in quark matter is quite different from
that in nuclear matter~\cite{Blaschkeflux,ABRflux}.
Nuclear matter is an electromagnetic superconductor
(because of proton-proton pairing which breaks
the $U(1)_Q$ gauge symmetry) and also a superfluid
(because of neutron-neutron pairing).
Magnetic fields are therefore restricted to Abrikosov flux tubes,
and angular momentum is carried by rotational vortices.
The magnetic
flux tubes can be dragged about by the outward motion of the rotational
vortices as the neutron star spins down  
\cite{Sauls,Dragging,Bhattacharya,Ruderman,RudermanTalk},
and can also be pushed outward if the gap at the 
proton fermi surface increases with depth within
the neutron star~\cite{HsuMag}.
One therefore expects the magnetic field of an isolated pulsar to
decay over billions of years as it spins down
\cite{Dragging,Bhattacharya,Ruderman,RudermanTalk}
or perhaps more quickly~\cite{HsuMag}.  
However, there is
no observational evidence for the decay of the magnetic
field of an isolated pular over periods of billions of
years~\cite{Bhattacharya,Lorimer}.

A color superconductor, on the other hand, leaves unbroken
a rotated electromagnetism $U(1)_\Qt$, a mixture of photon
and gluon, allowing long-range $\Qt$-magnetic fields. This is true of the
CFL phase, and also of the 2SC phase as long as the temperature
is high enough so that the blue
quarks do not pair.

The new unbroken rotated electromagnetic field $A_\Qt$ is
just a linear combination of the photon $A_\mu$ and one of the
gluons $G_\mu^8$,
\beq
A_\mu^\Qt= \cos\al_0 \,A_\mu + \sin\al_0\,G_\mu^8
\eeq
the orthogonal combination $A_\mu^{X}$ is massive.
The mixing angle $\alpha_0$ is the analogue of the Weinberg
angle in electroweak theory, in which the 
presence of the Higgs condensate causes the hypercharge
and $W_3$ gauge bosons to mix to form the photon, $A_\mu$, and 
the massive $Z$ boson.   
$\sin(\alpha_0)$ is proportional to $e/g$ and turns
out to be about $1/20$ in the 2SC phase and $1/40$ in the CFL
phase~\cite{ABRflux}.  This means that the 
$\Qt$-photon which propagates in color superconducting
quark matter is mostly photon with only 
a small gluon admixture. If a color superconducting neutron star core 
is subjected to an ordinary magnetic field, it will either
expel the $X$ component of the flux
or restrict it to flux tubes, but it 
admits the great majority of the flux
in the form of a $B_{\tilde Q}$ magnetic field satisfying
Maxwell's equations.   
The decay in time of this ``free field'' (i.e. not in flux tubes) 
is limited by the $\tilde Q$-conductivity of the quark matter.

The CFL phase contains no electrons, and all its charged modes
are gapped, making it an electromagnetic insulator.
The 2SC phase has electrons as well as blue quasiquarks,
and turns out to be a very good conductor.
Thus the 2SC and CFL phases, while both allowing
long-range $\Qt$-flux fields, react very differently to
attempts to {\em change} the magnetic field.
The CFL phase allows such changes, but the 2SC, as a near-perfect
conductor, generates eddy currents that oppose the change,
locking the magnetic field into the core with a decay time
of order $10^{13}$ years~\cite{ABRflux}

This means that a 2SC quark matter core within a neutron
star can act as an ``anchor'' for the magnetic field,
preventing the flux-tube-dragging mechanism
that can operate in ordinary nuclear matter.
Even though this distinction is a qualitative one, it
will be 
difficult to confront it with data since what is
observed is the total dipole moment of the neutron star.
A color superconducting
core can only anchor those magnetic flux lines which pass through
the core, while in a neutron star with no quark matter core
the entire internal magnetic field can decay over time. 
In both cases, however, the total dipole moment can change
since the magnetic flux lines which do not pass through
the core can move.

\subsection{Glitches and the crystalline color superconductor}

The crystalline
LOFF phase has been discussed above.  It occurs when two different
types of quark have different Fermi momenta (because their masses or
chemical potentials are different) and are just barely able to pair.

Such situations are likely to be generic in nature, where, because of
the strange quark mass, combined with requirements of weak equilibrium
and charge neutrality, all three flavors of quark in general have
different chemical potentials. 
To date the LOFF condensate has only been studied in simplified
two-flavor models, so it is not clear whether it can be expected
to occur in compact stars. However, in the model a LOFF phase occured
if the gap $\Delta_0$ which characterizes the uniform
color superconductor present at smaller values of $\delta\mu$ was
about 40~MeV~\cite{OurLOFF}. This is in the middle of the range of present
estimates of superconducting gaps. 
It is therefore worthwhile to consider the consequences.

\subsubsection{Glitches and vortex pinning}
Glitches are sudden
jumps in rotation frequency $\Omega$ of a pulsar, which may
be as large as $\Delta\Omega/\Omega\sim 10^{-6}$, but may also
be several orders of magnitude smaller. The frequency of observed
glitches is statistically consistent with the hypothesis that 
all radio pulsars experience glitches~\cite{AlparHo}.
Glitches are thought to originate in
the rigid neutron star crust, typically somewhat 
more than a kilometer thick, where rotational vortices in a
neutron superfluid are pinned to the crystal structure of the crust.
As the pulsar's spin gradually slows,
the vortices must gradually move outwards since the rotation frequency
of a superfluid is proportional to the density of vortices. 
Models~\cite{GlitchModels} differ
in important respects as to how the stress associated
with pinned vortices is released in a glitch: for example,
the vortices may break and rearrange the crust, or a cluster
of vortices may suddenly overcome the pinning force and 
move macroscopically outward, with
the sudden decrease in the angular momentum
of the superfluid within the crust resulting in a sudden increase
in angular momentum of the rigid crust itself and hence a glitch.
All the models agree that the fundamental requirements
are the presence of rotational vortices in a superfluid 
and the presence
of a rigid structure which impedes the motion of vortices and
which encompasses enough of the volume of the pulsar to contribute
significantly to the total moment of inertia.

Henceforth, we suppose  that the LOFF phase is a superfluid, 
which means that if it occurs within a pulsar it will be threaded
by an array of rotational vortices.
It is reasonable to expect that these vortices will
be pinned in a LOFF crystal, in which the
diquark condensate varies periodically in space.
Indeed, one of the suggestions for how to look for a LOFF phase in
terrestrial electron superconductors relies on the fact that
the pinning of magnetic flux tubes (which, like the rotational vortices
of interest to us, have normal cores)
is expected to be much stronger
in a LOFF phase than in a uniform BCS superconductor~\cite{Modler}.
Note that the chiral crystal phase~\cite{RappCrystal} is
not a superfluid, so it will not contain rotational vortices.

\subsubsection{Vortex pinning in the LOFF phase}
A real calculation of the pinning force experienced by a vortex in a
crystalline color superconductor must await the determination of the
crystal structure of the LOFF phase. We can, however, attempt an order
of magnitude estimate along the same lines as that done by Anderson
and Itoh~\cite{AndersonItoh} for neutron vortices in the inner crust
of a neutron star. In that context, this estimate has since been made
quantitative~\cite{Alpar77,AAPS3,GlitchModels}.  
For one specific choice of parameters~\cite{OurLOFF}, 
the resulting pinning force per unit length of vortex was estimated
essentially by dimensional analysis at
\beq
\mbox{LOFF:}\quad f_p  \sim  (4 {\rm \ MeV})/(80 {\rm \ fm}^2).
\eeq
It is premature to compare such a crude result 
to the results of serious calculations 
\cite{Alpar77,AAPS3,GlitchModels}, but it is
remarkable that they prove to be similar: 
the pinning force per unit length for neutron vortices
in the inner crust is
\beq
\mbox{neutron star:}\quad f_p\approx(1-3 {\rm ~MeV})/(200-400 {\rm ~fm}^2).
\eeq

This raises the possibility that pulsars might be strange stars after all
\cite{HZS,AFO}.
Strange quark stars are made almost
entirely of quark
matter with either no hadronic matter content at all or perhaps
a thin crust, of order one hundred meters thick, which contains
no neutron superfluid~\cite{AFO,GlendenningWeber}.
No successful models of glitches in the crust
of a strange quark star have been proposed, indicating that
pulsars are not strange stars~\cite{Alpar,OldMadsen,Caldwell}.
The possibility of a shell of crystalline LOFF quark matter
inside a quark star revives the possibility that glitches
could occur in quark stars, as a result of the pinning
of quark-superfluid vortices to the LOFF crystal.

\section{CONCLUSIONS}

The possibility of quark pairing and color superconductivity
in high-density QCD
in an intriguing one. We have learnt much about the
rich structure of phases that it leads to, as
outlined in sections \ref{sec:2flavor}
to \ref{sec:unlock} above.
There are many directions that remain to be
explored, from new pairing structures to
detailed studies of the known 2SC, 2SC+s, CFL, and LOFF phases.
It has even been suggested that zero-density
QCD can be understood in terms of a quark-paired condensate
in combination with an adjoint chiral condensate~\cite{BW}.

The most pressing task, however,
is to investigate the consequences of
our findings for the phenomenology of neutron/quark stars, which are
the only naturally occuring example of cold matter at the densities we
have studied.  The first steps in this directions have been 
described in section \ref{sec:compact}.
Much theoretical work remains to be done before
we can make sharp proposals for 
astrophysical observations that might teach us 
whether compact stars contain quark matter, and the nature
of the quark pairing.

\vspace{3ex}
\noindent ACKNOWLEDGMENTS\\
I thank my collaborators J. Berges, J. Bowers,
K. Rajagopal, F. Wilczek, and also
S. Hsu, J. Madsen, S. Reddy, T. Sch\"afer for discussions
and communications.


\begin{thebibliography}{99}

\bibitem{Karsch_Tc}
Karsch F. {\it Nucl. Phys. (Proc. Suppl.)} 83-84:14 (2000)

\bibitem{Reviews}
Hong, D. hep-ph/0101025 (2001);
Rajagopal K, Wilczek F. hep-ph/0011333 (2000);
Sch\"afer T, Shuryak E. hep-ph/0010049 (2000);
Rajagopal K. hep-ph/0009058 (2000);
Rischke D, Pisarski R. 
  Proceedings of the "Fifth Workshop on QCD", Villefranche, Jan. 3-7, 2000;
Hsu S. hep-ph/0003140 (2000);
Alford M. hep-ph/0003185 (2000);

\bibitem{BCS}
Bardeen J, Cooper LN, Schrieffer JR. {\it Phys. Rev.} 106:162  (1957);
{\it Phys. Rev.} 108:1175  (1957)

\bibitem{Barrois}
Barrois B, {\it Nucl. Phys.} B129:390  (1977); 
Frautschi S. 
"Proceedings of workshop on hadronic matter at extreme density",
Erice 1978

\bibitem{BarroisPhD}
Barrois B. ``Nonperturbative effects in dense quark matter'',
Cal Tech PhD thesis, UMI 79-04847-mc (1979)

\bibitem{BailinLove}
Bailin D, Love A. {\it Phys. Rept.} 107:325  (1984), and references therein

\bibitem{ARW2}
Alford M, Rajagopal K, Wilczek F. {\it Phys. Lett.} B422:247  (1998)

\bibitem{RappETC}
Rapp R, Sch\"afer T, Shuryak EV, Velkovsky M. 
  {\it Phys. Rev. Lett.} 81:53  (1998)

\bibitem{Son}
Son DT. {\it Phys. Rev.} D59:094019  (1999)

\bibitem{SW-pert} 
Sch\"afer T, Wilczek F. {\it Phys. Rev.} D60:114033  (1999)

\bibitem{PR-pert} 
Pisarski RD, Rischke DH. {\it Phys. Rev.} D61:074017  (2000)

\bibitem{Hong-pert}
Hong, DK. {\it Nucl. Phys.} B 582:451 (2000); 
Hong, DK. {\it Phys. Lett.} B473:118 (2000)


\bibitem{PR-Tc} 
Pisarski RD, Rischke DH. {\it Phys. Rev.} D61:051501  (2000)

\bibitem{HMSW} 
Hong DK, Miransky VA, Shovkovy IA, Wijewardhana LC.
  {\it Phys. Rev.} D61:056001  (2000); 
erratum {\it Phys. Rev.} D62:059903 (2000)

\bibitem{rockefeller} 
Brown WE, Liu JT, Ren H. {\it Phys. Rev.} D61:114012  (2000); 
{\it Phys. Rev.} D62:054016  (2000); 
{\it Phys. Rev.} D62:054013  (2000)

\bibitem{Hsu2} 
Hsu SD, Schwetz M. {\it Nucl. Phys.} B572:211  (2000)

\bibitem{SchaeferPatterns} 
Sch\"afer T. {\it Nucl. Phys.} B575:269  (2000)

\bibitem{ShovWij} 
Shovkovy IA, Wijewardhana LC. {\it Phys. Lett.} B470:189  (1999)

\bibitem{RajagopalShuster}
Rajagopal K, Shuster E. hep-ph/0004074 (2000)

\bibitem{AKS}  
Agasian N, Kerbikov B, Shevchenko, V.
{\it Phys. Rept.}  320:131 (1999) 

\bibitem{EKT}
Ebert D, Klimenko K, Toki H. hep-ph/0011273

\bibitem{BergesRajagopal}
Berges J, Rajagopal K. {\it Nucl. Phys.} B538:215  (1999)

\bibitem{CarterDiakonov}
Carter G W, Diakonov D. {\it Phys. Rev.} D60:016004  (1999)

\bibitem{IwaIwa}
Iwasaki M, Iwado T. {\it Phys. Lett.} B350:163  (1995);
Iwasaki M. {\it Prog. Theor. Phys. Suppl.} 120:187 (1995);

\bibitem{ARW3}
Alford M, Rajagopal K, Wilczek F, {\it Nucl. Phys.} B537:443  (1999)

\bibitem{SW-RG}
Sch\"afer T, Wilczek F. {\it Phys. Lett.} B450:325  (1999)

\bibitem{EHS}
Evans N, Hsu S, Schwetz M, hep-ph/9810514, hep-ph/9808444

\bibitem{randmat}
Vanderheyden B, Jackson AD. {\it Phys. Rev. D62:094010} (2000);
Pepin S, Sch\"afer A. hep-ph/0010225 (2000)

\bibitem{SU2unbroken} Rischke D, Son D, Stephanov M. hep-ph/0011379 (2000)

\bibitem{1SC}
Alford M, Bowers J, Rajagopal K, in preparation

\bibitem{Sannino} Sannino F. {\it Phys. Lett.} B480:280 (2000)

\bibitem{SanHsu} Hsu S, Sannino F, Schwetz M.
hep-ph/0006059 (2000)

\bibitem{Berges}
Berges J. hep-ph/0012013 (2000)


\bibitem{SS}
Srednicki M, Susskind L. {\it Nucl. Phys.} B187:93 (1981)


\bibitem{SW-cont}
Sch\"afer T, Wilczek F. {\it Phys. Rev. Lett.} 82:3956  (1999)

\bibitem{HsuCFL}
Evans N, Hormuzdiar J, Hsu S, Schwetz M.
  {\it Nucl. Phys.} B581:391  (2000)

\bibitem{PisarskiCFL}
Pisarski R, Rischke D. Proceedings of the Judah Eisenberg Memorial Symposium,
 'Nuclear Matter, Hot and Cold', Tel Aviv, April 14 - 16, 1999

\bibitem{ABR2+1}
Alford M, Berges J, Rajagopal K. {\it Nucl. Phys.} B558:219  (1999)

\bibitem{Halasz} Halasz M, Jackson A, Shrock R, Stephanov M,
Verbaarschot J. {\it Phys. Rev.} D58:096007  (1998)

\bibitem{BJW2}
Berges J, Jungnickel D-U, Wetterich C. hep-ph/9811387

\bibitem{PisarskiRischke1OPT}
Pisarski RD, Rischke DH. {\it Phys. Rev. Lett.} 83:37  (1999)

\bibitem{StephanovRandMat}
Stephanov M. {\it Phys. Rev. Lett.} 76:4472 (1996)

\bibitem{Schaefer1Flavor}
Sch\"afer T. hep-ph/0006034 (2000)

\bibitem{KaplanNelson}
Kaplan DB, Nelson AE. {\it Phys. Lett.} B175:57  (1986)

\bibitem{BrownRho}
Brown G, Kubodera K, Rho M. {\it Phys. Lett.} B175:57 (1987)

\bibitem{SchaeferKaon}
Sch\"afer T. hep-ph/0007021 (2000)

\bibitem{CFLneutral}
Rajagopal K, Wilczek F. hep-ph/0012039 (2000)

\bibitem{Pisarski} 
Pisarski R. {\it Phys. Rev.} C62:035202 (2000)

\bibitem{Iwa3flav}
Iwasaki M, Iwado T. {\it Prog. Theor. Phys.} 94:1073 (1995)

\bibitem{RappCrystal}
Rapp R, Shuryak E, Zahed I. hep-ph/0008207 (2000)

\bibitem{Clogston}
Clogston AM. {\it Phys. Rev. Lett.} 9:266  (1962); 
Chandrasekhar BS. {\it App. Phys. Lett.} 1:7  (1962)

\bibitem{OurLOFF}
Alford M, Bowers J, Rajagopal K, hep-ph/0008208 (2000)

\bibitem{Sedrakian}
Sedrakian A, Lombardo U. {\it Phys. Rev. Lett.} 84:602  (2000)

\bibitem{SW2}
Sch\"afer T, Wilczek F. {\it Phys. Rev.} D60:074014  (1999)

\bibitem{Bedaque}
Bedaque PF. hep-ph/9910247 (1999)

\bibitem{LO}
Larkin AI, Ovchinnikov YuN. {\it Zh. Eksp. Teor. Fiz.} 47:1136 (1964); 
translation: {\it Sov. Phys. JETP} 20:762  (1965)

\bibitem{FF}
Fulde P, Ferrell RA. {\it Phys. Rev.} 135:A550 (1964)

\bibitem{Takada1}
Takada S, Izuyama T. {\it Prog. Theor. Phys.} 41:635  (1969)

\bibitem{HP-nstar}
Heiselberg H, Pandharipande V. 
  {\it Annu. Rev. Nucl. Part. Sci.} 50:481 (2000)

\bibitem{Page}
Page D, Prakash M, Lattimer JM, Steiner A. hep-ph/0005094 (2000)

\bibitem{Schaab}
Schaab C, et al. {\it Astrophys. Lett. J.} 480:L111  (1997)
and references therein

\bibitem{Blaschke}
Blaschke D, Klahn T, Voskresensky DN. 
  {\it Astrophys. J.} 533:406 (2000)

\bibitem{effth} 
Son D, Stephanov M. {\it Phys. Rev.} D61:074012 (2000);
Rho M, Wirzba A, Zahed I.  {\it Phys. Lett.} B473:126 (2000);
Casalbuoni R, Gatto R.  hep-ph/9911223 (1999); 
Hong D, Lee T, Min D.  {\it Phys. Lett.} B477:137 (2000);
Manuel C, Tytgat M.  {\it Phys. Lett.} B479:190 (2000);
Beane S, Bedaque P, Savage M. {\it Phys. Lett.} B483:131 (2000)

\bibitem{CarterReddy}
Carter G W, Reddy S. hep-ph/0005228 (2000)

\bibitem{Madsen}
Madsen J. {\it Phys. Rev. Lett.} 85:10  (2000)

\bibitem{Bildsten}
Bildsten L, Ushomirsky G, astro-ph/9911155 (1999)

\bibitem{Blaschkeflux}
Blaschke D, Sedrakian DM, Shahabasian KM. 
{\it Astron. Astrophys.} 350:L47 (1999)

\bibitem{ABRflux}
Alford M, Berges J, Rajagopal K. {\it Nucl. Phys.} B571:269  (2000)

\bibitem{Sauls}
Sauls J. in {\it Timing Neutron Stars}, \"Ogleman J and van den 
Heuvel EPJ, eds., (Kluwer, Dordrecht: 1989) 457. 

\bibitem{Dragging}
Srinivasan G, Bhattacharya D,
Muslimov AG, Tsyagan AI. {\it Curr. Sci.} 51:31 (1990)

\bibitem{Bhattacharya}
Bhattacharya D, Srinivasan G. {\it X-Ray Binaries}, 
Lewin WHG, van Paradijs J, van den Heuvel EPJ eds.,
(Cambridge University Press, 1995) 495

\bibitem{Ruderman}
Ruderman M. {\it Astrophys. J.} 366:261 (1991); 
{\it Astrophys. J.} 382:576 (1991);
{\it Astrophys. J.} 382:587 (1991)

\bibitem{RudermanTalk}
Ruderman M, Zhu T, Chen K. {\it Astrophys. J.} 492:267 (1998)

\bibitem{HsuMag}
Hsu S. nucl-th/9903039 (1999)

\bibitem{Lorimer}
Lorimer D, Bailes M, Harrison P. {\it MNRAS} 289:592 (1997)

\bibitem{AlparHo}
Alpar MA, Ho C. {\it Mon. Not. Astron. Soc. R.} 204:655  (1983).
For a recent review, see Lyne, A in {\it Pulsars: Problems and Progress},
Johnston S, Walker MA, Bailes M, eds., 73 (ASP, 1996)

\bibitem{GlitchModels}
For reviews, see Pines D, Alpar A. {\it Nature} 316:27  (1985); 
Pines D, in {\it Neutron Stars: Theory and Observation}, 
  Venturaand J Pines D, eds., 57 (Kluwer, 1991); 
Alpar MA, in {\it The Lives of Neutron Stars}, 
Alpar MA et al., eds., 185 (Kluwer, 1995).
For more recent developments and references to further work,see
Ruderman M, {\it Astrophys. J.} 382:587  (1991); 
Epstein RI, Baym G, {\it Astrophys. J.} 387:276  (1992); 
Alpar MA, Chau HF, Cheng KS, Pines D, {\it Astrophys. J.} 409:345  (1993); 
Link B, Epstein RI, {\it Astrophys. J.} 457:844  (1996); 
Ruderman M, Zhu T, Chen K, {\it Astrophys. J.} 492:267  (1998); 
Sedrakian A, Cordes JM, {\it Mon. Not. Astron. Soc. R.} 307:365  (1999)

\bibitem{Modler}
Modler R, et al. {\it Phys. Rev. Lett.} 76:1292  (1996)

\bibitem{AndersonItoh}
Anderson PW, Itoh N. {\it Nature} 256:25  (1975)

\bibitem{Alpar77}
Alpar MA, {\it Astrophys. J.} 213:527  (1977)

\bibitem{AAPS3}
Alpar MA, Anderson PW, Pines D, Shaham J. {\it Astrophys. J.} 278:791  (1984)

\bibitem{HZS}
Haensel P, Zdunik JL, Schaeffer R. {\it Astron. Astrophys.} 160:121  (1986)

\bibitem{AFO}
Alcock C, Farhi E, Olinto A. {\it Phys. Rev. Lett.} 57:2088  (1986); 
{\it Astrophys. J.} 310:261  (1986)

\bibitem{GlendenningWeber}
Glendenning NK, Weber F. {\it Astrophys. J.} 400:647  (1992)

\bibitem{Alpar}
Alpar A. {\it Phys. Rev. Lett.} 58:2152  (1987)

\bibitem{OldMadsen}
Madsen J. {\it Phys. Rev. Lett.} 61:2909  (1988)

\bibitem{Caldwell}
Caldwell RR, Friedman JL. {\it Phys. Lett.} B264:143  (1991)

\bibitem{BW}
Berges J, Wetterich C.  hep-ph/0012311 (2000)


\end{thebibliography}
\end{document}